\documentstyle[12pt]{article}
\makeatletter \oddsidemargin  -.1in \evensidemargin -.1in
\newcommand{\singlespacing}{\let\CS=\@currsize\renewcommand{\baselinestretch}{1}\tiny\CS}
\newcommand{\oneandahalfspacing}{\let\CS=\@currsize\renewcommand{\baselinestretch}{1.25}\tiny\CS}
\newcommand{\doublespacing}{\let\CS=\@currsize\renewcommand{\baselinestretch}{1.35}\tiny\CS}

\newtheorem{rule-def}[theorem]{Rule}

\RequirePackage[dvips]{graphicx} \textheight 22.5cm

\begin{document}

%\singlespacing
%\doublespacing
\title {\bf Peristaltic Transport of a Couple Stress Fluid : Some Applications to Hemodynamics}
\author{\small S. Maiti$^1$\thanks{Email address: {\it somnathm@cts.iitkgp.ernet.in (S. Maiti)}},~~~J.C.Misra$^2$\thanks{Email address: {\it misrajc@rediffmail.com (J. C. Misra)}}~  \\
$^1$\it School of Medical Science and Technology $\&$ Center for Theoretical Studies, \\IIT, Kharagpur-721302, India \\ \it$^2$Professor of Applied of Mathematics and Pro-Vice-chancellor (Academic),\\ SOA University, Bhubaneswar-751030, India\\ \it (Formerly, Professor and Chairman, Department of Mathematics,\\ \it IIT Kharagpur, India)\\}
\date{}
\maketitle \noindent \doublespacing
\begin{abstract}
The paper deals with a theoretical investigation of the
peristaltic transport of a couple stress fluid in a porous
channel. The study is motivated towards investigating the
physiological flow of blood in the micro-circulatory system, by taking
account of the particle size effect. The velocity, pressure gradient,
stream function and frictional force of blood are investigated, when
the Reynolds number is small and the wavelength is large, by using
appropriate analytical and numerical methods. Effects of different
physical parameters reflecting porosity, Darcy number, couple stress
parameter as well as amplitude ratio on velocity profiles, pumping
action and frictional force, streamlines pattern and trapping of blood
are studied with particular emphasis. The computational results are
presented in graphical form. The results are found to be in good
agreement with those of Shapiro et. al \cite{shapiro} that was carried
out for a non-porous channel without consideration of couple stress
effect. The present study puts forward an important observation that
for peristaltic transport of a couple stress fluid during free
pumping, flow reversal can be considerably controlled by suitably
adjusting the couple stress effect of the fluid/Darcy permeability of
the channel. It is also possible to avoid the occurrence of trapping,
by reducing the permeability.\\

\end{abstract}

\section{Introduction}
Peristalsis is a natural mechanism of pumping that is observed in the
case of most physiological fluids. In the transport of some other
fluids also peristaltic behaviour is observed. This behaviour is
usually associated with a progressive wave of area contraction or
expansion along the length of the boundary of a fluid-filled
distensible tube. Peristaltic transport occurs in various
physiological activities, for example, in the flow of urine from
kidney to the bladder, in the movement of food material through the
digestive tract, in flow of fluids through lymphatic vessels as well as in
semen movement in the vas differences, in the movement of bile from
gall bladder into the duodenum and spermatozoa inside the ductus
efferentes of the male reproductive tract and cervical canal, in
flow of ovum in the fallopian tube, the movement of cilia and also in
the flow of blood through small blood vessels. This phenomenon is also
applied in the propulsion of some industrial fluids.
\begin{center}
\begin{tabular}{|l l|}\hline
{~\bf Nomenclature} &~ \\
~~$a$ & Wave amplitudes\\
~~$d $ & Half-width of the channel \\
~~$Da $ & Darcy number \\
~~$H $ & Vertical displacement of the wall \\
~~$k $ & Permeability parameter \\
~~$\alpha, \eta$ & Couple stress parameters \\
~~$P $ & Fluid pressure\\
~~$Q $ & Flux at any axial station\\
~~$R $ & Reynolds number of the fluid\\
~~$t $  & Time\\
~~$X,Y $ & Rectangular Cartesian co-ordinates\\
~~$U,V$ & Velocity components in X,Y directions respectively\\
~~$m_1, m_2$ & Constants defined in equation (18)\\
~~$\delta $ & Wave number\\
~~$\Delta p$ & Pressure difference between the channel ends\\
~~$\epsilon $ & Porosity parameter\\
~~$\lambda $ & Wave length of the traveling wave motion of the wall\\
~~$\mu $ & Viscosity of the fluid\\
~~$\nu $ & Kinematic viscosity of the fluid\\
~~$\phi$ & Amplitude ratio \\
~~$\rho$ & Density of the fluid \\
\hline
\end{tabular}
\end{center}
This mechanism is also used in many biomedical appliances, such as
finger pumps, heart-lung machine, blood pump machine, dialysis machine
and also in industries for the transport of noxious fluid in nuclear
industries, as well as in roller pumps. For this reason in current
years, studies of peristaltic movement have been receiving growing
interest of scientific researchers. Much of the early literature of
theoretical investigations, arranged according to the geometry, the
type of fluids, the Reynolds number, wave amplitude, wavelength and
the wave shape, along with an account of experimental studies on
peristaltic transport have been reviewed by Srivastava and Srivastava
\cite{srivastava1}. Some of the important theoretical studies on
peristalsis have been discussed by Eytan et al. \cite{eytan},
Jimenez-Lozano et al. \cite{jimenez-lozano}, Misra et al
\cite{misra1,misra2,misra3,misra4,misra5,misra6,misra7,misra8}, Maiti
and Misra \cite{maiti}, Usha and Rao \cite{usha}, Mishra and Rao
\cite{mishra1}, Rao and Mishra \cite{rao}, Park et al \cite{park} as
well as by Akbar and Nadeem\cite{akbar}. A few of these studies have
been carried out by using the lubrication theory by neglecting the
fluid inertia and wall curvature without any restriction of the wave
amplitude. Some other analyses are based on the consideration of small
peristaltic wave amplitude, where the Reynolds number is
arbitrary. Antanovskii and Ramkissoon \cite{antanovskii} studied the
peristaltic transport of a compressible viscous fluid through a pipe
with the help of lubrication theory when the pressure drop changes
with time, by taking into account the wall deformation of the pipe.

Bergel \cite{bergel} observed that the capillary walls are surrounded
by flattened endothelial cell layers which are porous. Dash et
al. \cite{dash} considered the Brinkman equation to model blood flow
in a coronary artery in the pathological state when clogging of blood
occurs in the lumen of the artery. They assumed the clogged region as
a porous medium and assumed the permeability to be constant/varying in
the radial direction. Misra et
al. \cite{misra9,misra10,misra11,misra12,misra13,misra14} studied
different aspects of blood flow in arteries in normal/pathological
states. Some models on blood rheology have been discussed by Chien and
Skalak (cf. \cite{chien,beg1}).

It is well known that most physiological fluids including blood behave
as a non-Newtonian fluid. It has been suggested that blood flow
behaviour in small vessels (of diameter less than 0.2 mm) at low shear
rate ($<20 sec^{-1}$) can be represented by a power law fluid
\cite{charm1,charm2}. It was reported by Blair and Spanner
\cite{blair} that blood can be represented by Casson's model for
moderate shear rate flows. The non-Newtonian behaviour of blood is
mainly due to the suspension of red blood cells in the plasma. When
neutrally buoyant corpuscles are contained in a fluid and there exists
a velocity gradient due to shearing stress, corpuscles have rotatory
motion. Furthermore, it is observed that corpuscles have spin angular
momentum, in addition to orbital angular momentum. As a result, the
symmetry of stress tensor is lost in the fluid motion that is
subjected to spin angular momentum. The fluid that has neutrally
buoyant corpuscles, when observed macroscopically, exhibits
non-Newtonian behaviour, and its constitutive equation is expressed by
a stress tensor\cite{stokes}. The radius of gyration of the corpuscles
for such fluids is different from that of the fluid particles. Their
difference produces couple stress in the fluid. It is known that such
a fluid has spin angular momentum in addition to the couple stress
effect. The importance of consideration of couple stress effects in
studies of physiological and some other fluids was indicated by Cowin
\cite{cowin} and Beg et al \cite{beg2,beg3}.

 Studies pertaining to the couple stress fluid behaviour are very
 useful, because such studies bear the potential to better explain the
 behaviour of rheologically complex fluids, such as liquid crystals,
 polymeric suspensions that has have long-chain molecules, lubrication
 as well as human/sub-human blood
 \cite{stokes,srivastava2,zueco1,ghosh}. Valanis and Sun
 \cite{valanis} as well as Popel et al.\cite{popel} remarked that
 couple stress fluids constitute a special class of non-Newtonian
 fluids that take account of the particle size.  The couple stress
 theory presented by Stokes \cite{stokes} is quite suitable to study
 blood flow in micro-vessels by taking the size of the erythrocytes
 into account. Shehawey and Mekheimer \cite{shehawey} analysed the
 flow of a couple stress fluid for any arbitrary Reynolds number and
 wave number, considering the wave amplitude to be small.

It may be mentioned that consideration of porosity is very much
necessary to properly explain the fluid dynamical process that occurs
in different parts of the body, such as vascular beds, lungs, kidneys
and tumorous vessels. The importance of consideration of Darcian drag
effects in studies related to flow of blood and some other fluids
through porous media has been discussed by different researchers
\cite{rashidi,zueco2,beg4,beg5,beg6,zueco3,beg7,beg8}. Moreover, in
many biomechanical studies, porosity of the media has significant
influence on the transport of fluids. This applies more particularly
to vessels impeded by clots and also to highly perfused skeletal
tissues, tumors and to soft connecting tissues \cite{bhargava}. Some
further discussion on vessel porosity and peristaltic flows of
physiological fluids is available in several scientific publications
\cite{shapiro,takabatake1,takabatake2,sugihara-seki,guyton,nield,barbee}.

Being motivated by the observations reported in the above mentioned
studies, we have undertaken here a study that concerns peristaltic
flow of blood in a porous bed. With an aim to take account of the
particle size effect, blood has been modelled as a couple-stress
fluid.  The problem has been analysed by using lubrication theory
\cite{shapiro}. The analysis and the results of the study are
particularly applicable to the peristaltic transport of blood in
coronary arteries of smaller dimensions in the pathological state,
when the lumen of the segment of a small blood vessel turns into a
porous medium due to the presence of numerous blood clots, or when
arterial clogging takes place by deposition of fatty plaques of
cholesterol in the arterial lumen, and also in cases where numerous
tumors are grown inside the lumen due to excessive cell division.

The results of the present study will serve as a reasonably good
estimate of various fluid mechanical parameters for peristaltic
transport of blood in small blood vessels in a pathological
state. Since flow behaviour in an axi-symmetric vessel resembles that
in a channel, we have studied here two-dimensional channel flow of the
fluid. The results of this study will be applicable to blood vessels
in the micro-circulatory system without any restriction.

The study bears promise of multi-fold important applications. In the
realm of physiological fluid dynamics, in many situations it is
required to have estimates of a variety of fluid mechanical variables
when some physiological fluid has to pass through porous structures,
particularly in pathological states. The study will also have an
important bearing in examining the flow in a vessel when the luminal
surface of endothelial layer is attached with glycocalyx, which
contains a series of micro-molecules and adsorbed plasma proteins. The
present study can also find important applications in artificial
bio-processors that have synthetic porous surfaces. Since blood
flowing in small vessels has got relevance with flow in porous media,
the study is applicable to blood flow in micro-vessels more
particularly in situations where couple stress effects of blood are
prominent. In addition to all these, the analysis presented here
should find applications in polymer industries, where it is often
required to examine the flow behaviour of different types of fluids
passing through porous structures.
\begin{minipage}{1.0\textwidth}
     \begin{center}
       \includegraphics[width=3.5in,height=2.3in]{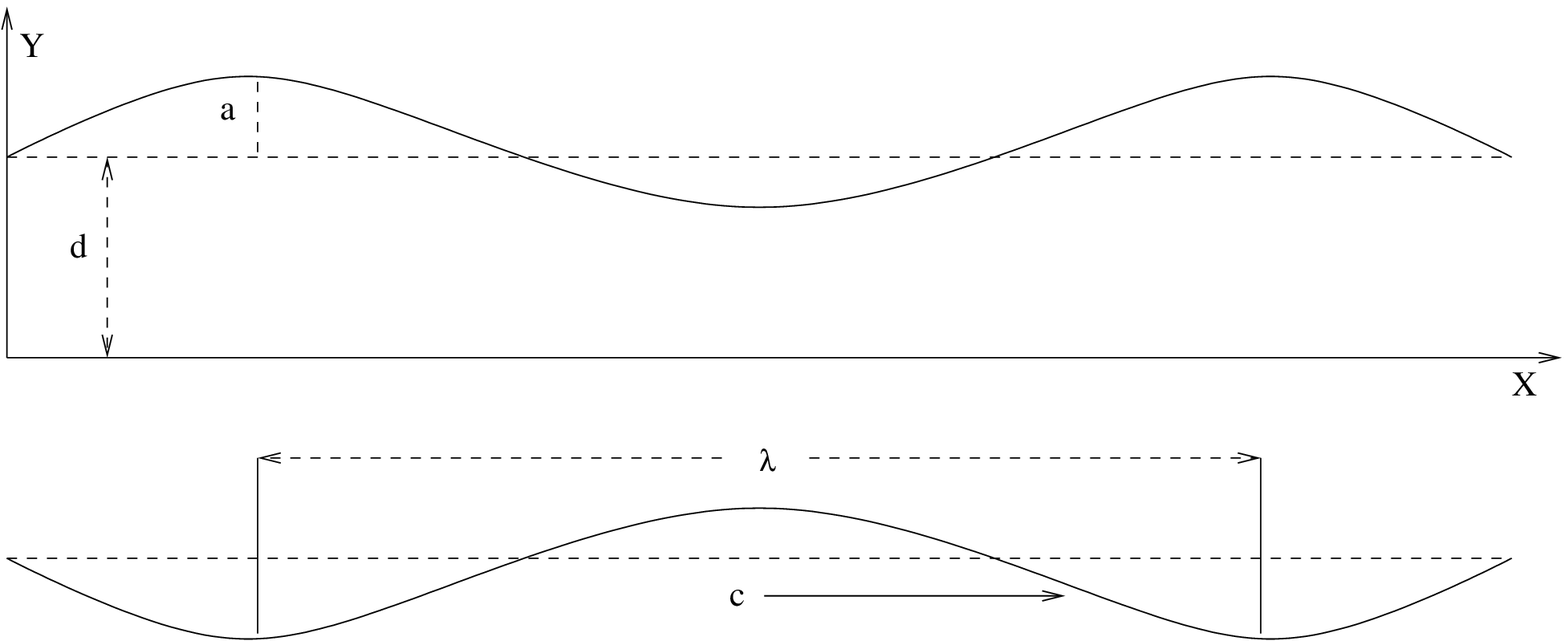}
\\ Fig. 1: A physical sketch of the problem   \\
\end{center}
\end{minipage}\vspace*{.25cm}

 \section{Mathematical Modelling and Analysis}
 Let us consider the peristaltic motion of blood on a porous
 channel. Blood is treated as a viscous couple stress fluid
 (non-Newtonian). We take (X,Y) as Cartesian coordinates of a point, X
 being measured in the direction of wave propagation and Y in the
 direction normal to the mean position of the corresponding small
 blood vessel. In the analysis that follows we consider channel flow.
 The constitutive equations and equations of motion for couple stress
 fluid flow in the absence of body moment and body couple can be put
 as \cite{stokes,srivastava2,zueco1,ghosh}
\begin{equation}
\tau_{ji,j}=\rho\frac{dv_i}{dt}
\end{equation}
\begin{equation}
e_{ijk}T_{jk}^A+M_{ji,j}=0
\end{equation}
\begin{equation}
\tau_{ij}=-P\delta_{ij}+2\mu d_{ij}
\end{equation}
\begin{equation}
\mu_{ij}=4\eta\omega_{j,i}+4\eta^\prime\omega_{i,j}
\end{equation}
in which $\rho$ is the fluid density, $\tau_{ij}$ and $T_{ij}^A$ designate
the symmetrical and antisymmetrical parts of the stress tensor $T_{ij}$,
$v_i$ the velocity vector, $M_{ij}$ the couple stress tensor,
$\mu_{ij}$ designate the deviatoric part of $M_{ij}$, $\omega_i$ stands for the
vorticity vector, $d_{ij}$ is the symmetric part of the velocity
gradient, $\eta$ and $\eta^\prime$ are constants associated with
the couple stress effect, P the pressure.

Let $ Y=H$ and $Y=-H$ be
 respectively the upper and lower boundaries of the channel
 (cf. Fig. 1). The motion is considered to be induced by a progressive
 sinusoidal wave train propagating with a constant speed $c$ along the
 channel wall, so that we have\\ $~~~~~~~~~~~~~~~~~~~~~~~~~~~~~~~~~~~~~~~~~~~~~~~H=d+a~cos(\frac{2\pi}{\lambda}(X-ct)),
 $ \\ where $a$ is the wave amplitude, $\lambda$ the wave length and
 $d$ the half width of the channel. \\ We shall use Brinkman model for
 investigating the flow through the porous medium. Thus the equations that govern
 the flow of the couple stress fluid in the porous channel may be
 written in the form
\begin{equation}
\frac{\partial U}{\partial X}+\frac{\partial V}{\partial Y}=0
\end{equation}
\begin{equation}
\rho \left (\frac{\partial U}{\partial t}+U\frac{\partial U}{\partial X}+V\frac{\partial U}{\partial Y}\right )=-\frac{\partial P}{\partial X}+\frac{\mu}{\epsilon~} \nabla^2 U-\eta \nabla^4 U -\frac{\mu}{k}U
\end{equation}
\begin{equation}
and~\rho\left (\frac{\partial V}{\partial t}+U\frac{\partial V}{\partial X}+V\frac{\partial V}{\partial Y}\right )=-\frac{\partial P}{\partial Y}+\frac{\mu}{\epsilon~} \nabla^2 V-\eta \nabla^4 V-\frac{\mu}{k}V
\end{equation}
in which k is the permeability parameter, $\eta$ the couple stress
parameter, $\epsilon$ stands for the porosity parameter and $\nabla^2=\frac{\partial^2 }{\partial X^2}+\frac{\partial^2 }{\partial Y^2}$~. \\
Considering a wave frame $(x,y)$ that moves with a velocity $c$ away from the fixed frame $(X,Y)$, let us use the transformations
\begin{equation}
x=X-ct, ~~y=Y, ~~ u=U-c, ~~ v=V, ~~ p(x,y)=P(X,Y,t)
\end{equation}
in which  $(u,v)$ and $(U,V)$ are the velocity components in the wave frame and the fixed frame respectively, $p$ and $P$ stand for pressure in wave and fixed frames of reference.\\
 Therefore the equation governing the  flow of the fluid which is steady in the wave frame of reference can be written in the form
\begin{equation}
\frac{\partial u}{\partial x}+\frac{\partial v}{\partial y}=0
\end{equation}
\begin{equation}
\rho \left (u\frac{\partial u}{\partial x}+v\frac{\partial u}{\partial y}\right )=-\frac{\partial p}{\partial x}+\frac{\mu}{\epsilon~} \nabla^2 u-\eta \nabla^4 u -\frac{\mu}{k}u-\frac{\mu c}{k}
\end{equation}
\begin{equation}
\rho\left (u\frac{\partial v}{\partial x}+v\frac{\partial v}{\partial y}\right )=-\frac{\partial p}{\partial y}+\frac{\mu}{\epsilon~} \nabla^2 v-\eta \nabla^4 v-\frac{\mu}{k}v
\end{equation}
Let us now introduce the following non-dimensional variables: 
\begin{eqnarray}
&&\bar{x}=\frac{x}{\lambda}, ~~\bar{y}=\frac{y}{d}, ~~\bar{u}=\frac{u}{c},
~~\bar{v}=\frac{v}{c\delta}, ~~ \delta=\frac{d}{\lambda}, ~~\bar{p}=\frac{d^2p}{\mu c\lambda},~~
\bar{t}=\frac{ct}{\lambda}, ~~h=\frac{H}{d},  ~~\phi=\frac{a}{d},\nonumber \\
 ~~&&\alpha^2=\frac{\eta/ \mu}{d^2}=\frac{l^2}{d^2}, ~~ R=\frac{\rho c d}{\mu }, ~~
Da=\frac{k}{d^2}.
\end{eqnarray}
Here $l^2$ is a material constant that has dimension of length
square. It can be identified with a property that depends on the size
of the fluid molecule (cf. Singh \cite{singh}). Dropping the bars
over the symbols, the equations governing the steady flow of the fluid
can be rewritten as
\begin{equation}
\delta \frac{\partial u}{\partial x}+\frac{\partial v}{\partial y}=0
\end{equation}
\begin{equation}
R\delta \left (u\frac{\partial u}{\partial x}+v\frac{\partial u}{\partial y}\right )=-\frac{\partial p}{\partial x}+\frac{1}{\epsilon~} \nabla^2 u-\alpha^2 \nabla^4 u -\frac{1}{Da}u-\frac{1}{Da}
\end{equation}
\begin{equation}
R\delta^3\left (u\frac{\partial v}{\partial x}+v\frac{\partial v}{\partial y}\right )=-\frac{\partial p}{\partial y}+\frac{\delta^2}{\epsilon~} \nabla^2 v-\alpha^2 \delta^2 \nabla^4 v-\frac{\delta^2}{Da}v
\end{equation}
\begin{eqnarray}
in~ which~~~\nabla^2\equiv\delta^2 \frac{\partial^2 }{\partial x^2}+\frac{\partial^2 }{\partial y^2}\end{eqnarray}
{\bf Boundary Conditions:}\\
Keeping in view the physical conditions of the problem, the boundary conditions in the fixed frame may be put mathematically as
\begin{eqnarray}
(i)~~~~ U_Y=0,~~U_{YYY}=0~~ at~~ Y=0~~~~~~~~~~~~~~~~~~~~~~~~~~~~~~~~~~~~~~~~~~~~~~~~~~~~~~~~~~~\nonumber \\
(ii)~~~~ U=0~~,~~V=\frac{\partial H}{\partial t}~~,~~-(V_{XX}-U_{YX})H_X+(V_{XY}-U_{YY})=0~~ at~~ Y=H \nonumber\\
\end{eqnarray}
In the wave frame, these boundary conditions in terms of the non-dimensional variables will be transformed to
\begin{eqnarray}
(i)~~~ u_y=0,~~u_{yyy}=0~~ at~~ y=0~~~~~~~~~~~~~~~~~~~~~~~~~~~~~~~~~~~~~~~~~~~~~~~~~~~~~~~~~~\nonumber \\
(ii)~~~ u=0,~~v=\frac{\partial h}{\partial t},~~-(\delta^4 v_{xx}-\delta^2 u_{yx})h_x+(\delta^2 v_{xy}-u_{yy})=0 ~~at~~ y=h ~~~\nonumber\\
 \end{eqnarray}
Applying long wave length approximation $(\delta \ll 1 )$, for small
Reynolds number (cf. Shapiro et al \cite{shapiro}), the equations (14) and (15) reduce to 
\begin{equation}
0=-\frac{\partial p}{\partial x}+\frac{1}{\epsilon~} \nabla_1^2 u-\alpha^2 \nabla_1^4 u -\frac{1}{Da}u-\frac{1}{Da}
\end{equation}
\begin{equation}
and~0=-\frac{\partial p}{\partial y},
\end{equation}
 \begin{eqnarray*}where ~~\nabla_1^2\equiv\frac{\partial^2 }{\partial y^2}\end{eqnarray*}
Thus the pressure turns out to be a function of x only. The solution of equation (19) satisfying the boundary conditions(18) is found in the form
\begin{equation}
  u=\frac{Da~m_2^2~ cosh (m_1 y)}{(m_2^2-m_1^2)cosh(m_1 h)}\frac{dp}{dx}-\frac{Da~m_1^2~ cosh (m_2 y)}{(m_2^2-m_1^2)cosh(m_2 h)}\frac{dp}{dx}-Da\frac{dp}{dx}-1
\end{equation}
where
 \begin{eqnarray}
 -h \le y \le h ,~m_1^2=\frac{1-\sqrt{1-4 \epsilon~^2\alpha^2/Da}}{2\epsilon~\alpha^2}~ and ~ m_2^2=\frac{1+\sqrt{1-4\epsilon~^2 \alpha^2/Da}}{2\epsilon~\alpha^2}
\end{eqnarray}
The rate of volume flow 'q' through each section is a constant
(independent of both x and t). It is given by 
\[ q=\int_{0}^{h}u~dy. \]
Substituting (21) and performing the integration, we find 
\begin{equation}
q=\frac{Da~m_2^2~ tanh (m_1 h)}{m_1(m_2^2-m_1^2)}\frac{dp}{dx}-\frac{Da~m_1^2~ tanh (m_2 h)}{m_2(m_2^2-m_1^2)}\frac{dp}{dx}-Da~h\frac{dp}{dx}-h
\end{equation}
Hence the flux at any axial station in the fixed frame is found to be given by
\[ Q=\int_{0}^{h}(u+1)dy =q+h~, \]
while the expression for the time-averaged volumetric flow rate over one period $T(=\frac{\lambda }{c} $) of the peristaltic wave is obtained as
\[ \bar{Q}=\frac{1}{T}\int_{0}^{T}Qdt=\frac{1}{T}\int_{0}^{T}(q+h)dt=q+1~. \]
The pressure gradient obtained from equation (23) can be expressed as
\[ \frac{dp}{dx}=\frac{(q+h) m_1 m_2 (m_2^2-m_1^2)}{Da~m_2^3~ tanh(m_1h)-Da~m_1^3~ tanh(m_2h)-Da~hm_1 m_2 (m_2^2-m_1^2)}. \]
The pressure rise per wavelength and the frictional force can be calculated by using the relations
\[\Delta p=\int_{0}^{\lambda}\left(\frac{dp}{dx}\right)dx.\]
\[F=-\int_{0}^{\lambda}h\left(\frac{dp}{dx}\right)dx.\]
Integrating over one wave length, we find
\[\Delta p=\frac{m_1 m_2 (m_2^2-m_1^2)}{Da} \bar{Q}I_1+\frac{m_1 m_2 (m_2^2-m_1^2)}{Da}I_2 \]
\[F=\frac{m_1 m_2 (m_2^2-m_1^2)}{Da} \bar{Q}I_3+\frac{m_1 m_2 (m_2^2-m_1^2)}{Da}I_4 \]
where
\[ I_1=\int_{0}^{1}\frac{dx}{m_2^3~tanh(m_1h)-m_1^3~tanh(m_2h)- hm_1 m_2 (m_2^2-m_1^2)}~~ ;\]
\[ I_2=\int_{0}^{1}\frac{(h-1)dx}{m_2^3~tanh(m_1h)-m_1^3~tanh(m_2h)- hm_1 m_2 (m_2^2-m_1^2)} \]
\[ I_3=-\int_{0}^{1}\frac{h dx}{m_2^3~tanh(m_1h)-m_1^3~tanh(m_2h)- hm_1 m_2 (m_2^2-m_1^2)}~~ ;\]
\[ and~I_4=-\int_{0}^{1}\frac{h(h-1)dx}{m_2^3~tanh(m_1h)-m_1^3~tanh(m_2h)- hm_1 m_2 (m_2^2-m_1^2)}. \]
In the fixed frame of reference, the expression for the non-dimensional axial velocity reads
\begin{eqnarray*}
U(X,Y,t)=\frac{m_1 m_2 (\bar{Q}+h-1)}{m_2^3~tanh(m_1h)-m_1^3~tanh(m_2h)- hm_1 m_2 (m_2^2-m_1^2)}\times ~~\nonumber~~~~~~~~~\\~~~\left (\frac{ m_2^2~ cosh (m_1 Y)}{cosh(m_1 h)}-\frac{ m_1^2~ cosh (m_2 Y)}{cosh(m_2 h)}+m_1^2-m_2^2\right )~~~~~~~~~~~~~~~~~~~~~
\end{eqnarray*}
\[with~~h=1+\phi~cos[2\pi(X-t)]~~.~~~~~~~~~~ ~~~~~~~\]

\begin{minipage}{1.0\textwidth}
     \begin{center}
       \includegraphics[width=3.5in,height=2.3in]{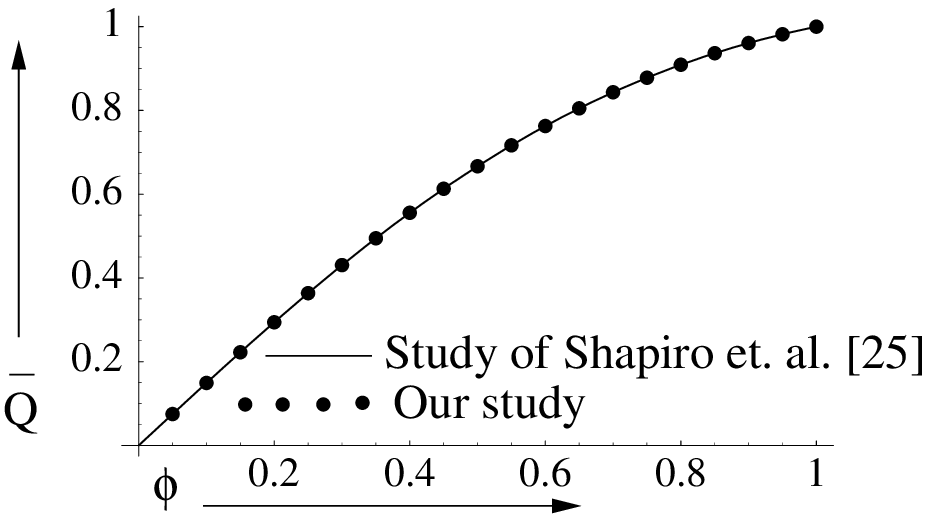}
\\Fig. 2.1: Variation of $\bar{Q}$ with  $\phi$ for $\Delta p$=0,~$\alpha=0.001$, Da=10000, e=0.9999.   \\
\end{center}
\end{minipage}\vspace*{.25cm}

\section{Results and Discussion}

Analytical expressions of the pressure gradient, the volumetric flow
rate, force of friction and axial velocity for the problem under consideration have been calculated and presented in the previous section. For the complexity of the problem, it was not possible to determine the analytical expression of $\Delta p$ and F in terms of  $\bar{Q}$ and other parameters. Therefore, we had to employ appropriate numerical methods along with the use of the software mathematica. Similarly, an analytical treatment was found inadequate for finding the velocity as a function of $\Delta p$, X, Y, t and other related parameters. This necessitated calculating first
\[\bar{Q}=\frac{Da\Delta p -m_1 m_2 (m_2^2-m_1^2)I_2}{m_1 m_2 (m_2^2-m_1^2)I_1}.\]

\begin{figure}
\fbox{\includegraphics[width=3.2in,height=2.3in]{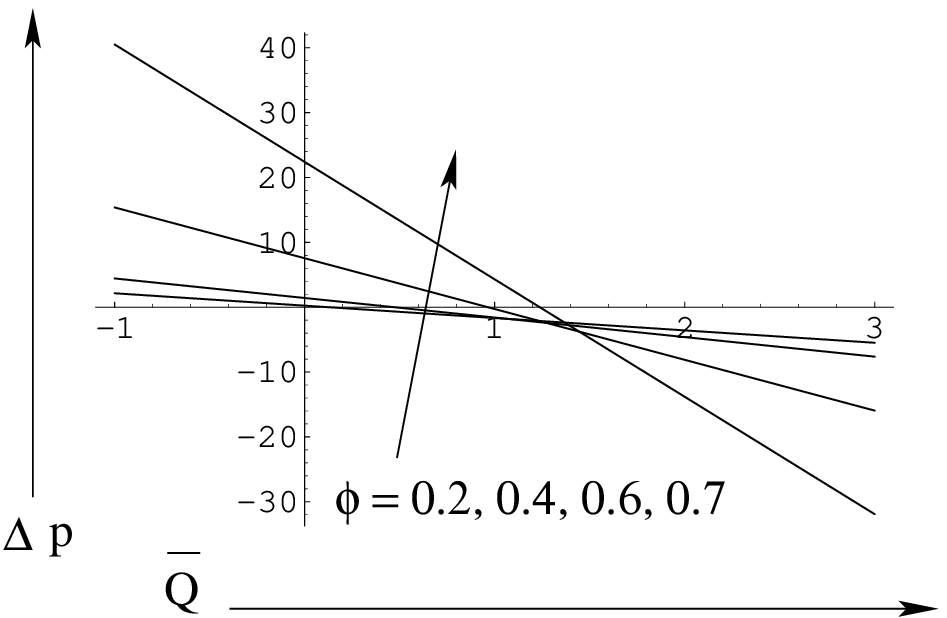}}\fbox{\includegraphics[width=3.2in,height=2.3in]{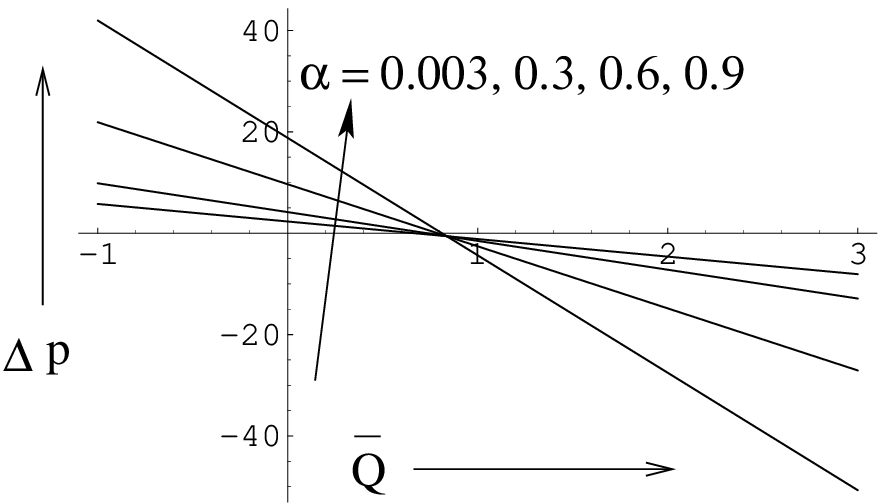}}
\\$~~~~~~~~~~~~~~~~~~~~~~Fig.~2.2~~~~~~~~~~~~~~~~~~~~~~~~~~~~~~~~~~~~~~~~~~~~~~~~~~~~~Fig.~2.3~~~~~~~~~~~~~~~~~~~~~~~~~$\\
\fbox{\includegraphics[width=3.2in,height=2.3in]{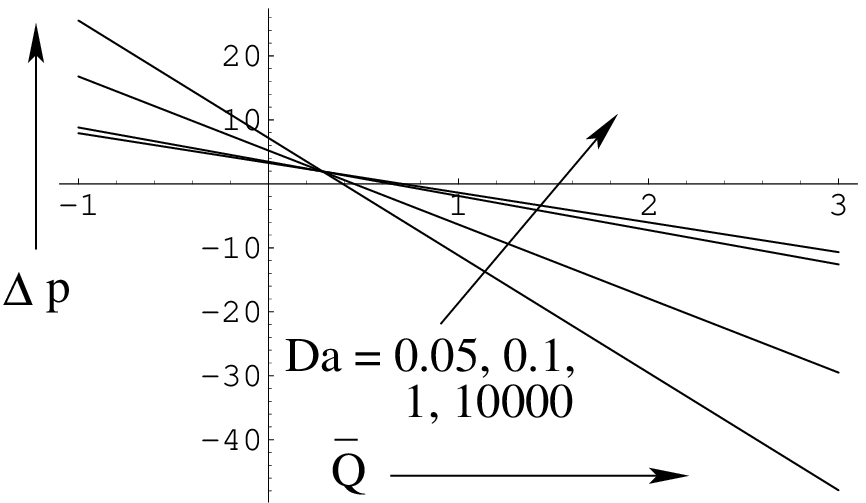}}\fbox{\includegraphics[width=3.2in,height=2.3in]{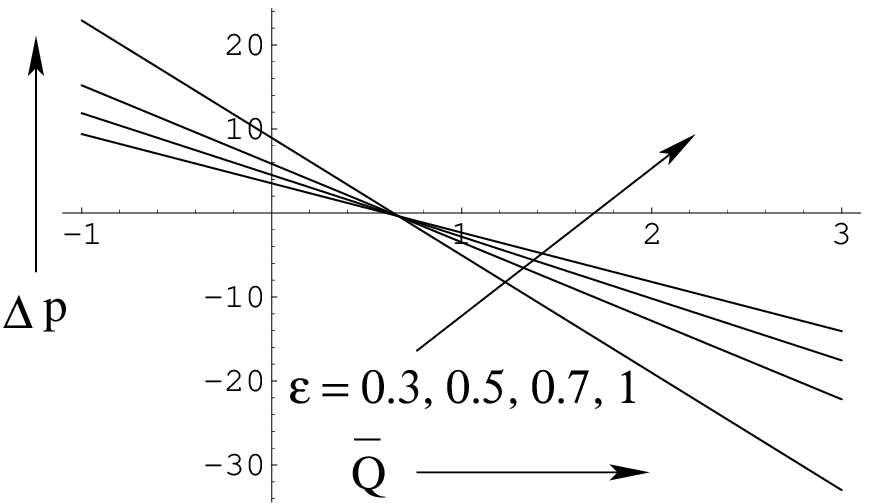}}
\\$~~~~~~~~~~~~~~~~~~~~~~Fig.~2.4~~~~~~~~~~~~~~~~~~~~~~~~~~~~~~~~~~~~~~~~~~~~~~~~~~~~~Fig.~2.5~~~~~~~~~~~~~~~~~~~~~~~~~~~$\\
Fig 2.2-2.5: Variation of $\bar{Q}$ with  $\Delta p$ (Fig 2.2) for different values of $\phi$ along with $\alpha$=0.2,~Da=10000, $\epsilon$=1; (Fig 2.3)  for different values of $\alpha$ with $\phi$=0.5, Da=10000, $\epsilon$=1; (Fig 2.4) for different values of Da with $\phi$=0.5, $\alpha$=0.2, $\epsilon$=0.95; (Fig 2.5) for different values of $\epsilon$ with $\phi$=0.5, $\alpha$=0.2, Da=0.5\\
\end{figure}

For the computation of this also, we had to make use of numerical quadratures. In this section, the problem will be investigated further for a specific situation by employing appropriate numerical integration techniques with the help of mathematica software. For the quantitative study, the following ranges of values for different parameters involved in the model analysis have been considered : \\
$\phi=$ 0.1~ to~ 0.9; $\Delta p =-40~to~40$;~ $\alpha=0.001$~ to ~1.0;
Da=0.05~ to~ 10000, e=0.3 to 1.0, $\bar{Q}=$~-1~to~3
(Cf. Refs.\cite{dash,guyton,nield,barbee}).

\begin{figure}
\includegraphics[width=3.3in,height=2.5in]{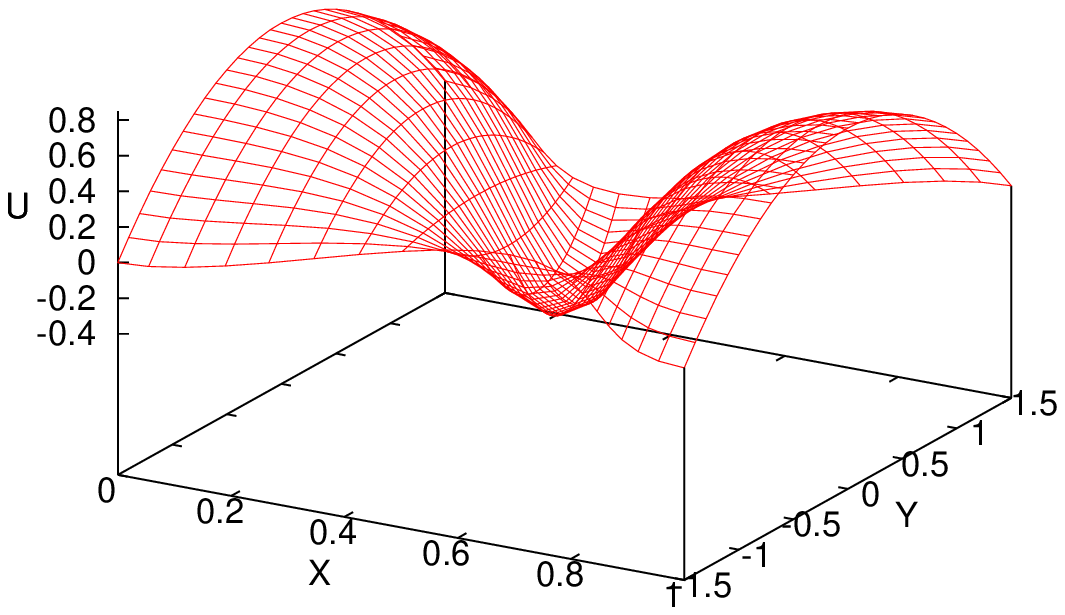}\includegraphics[width=3.3in,height=2.5in]{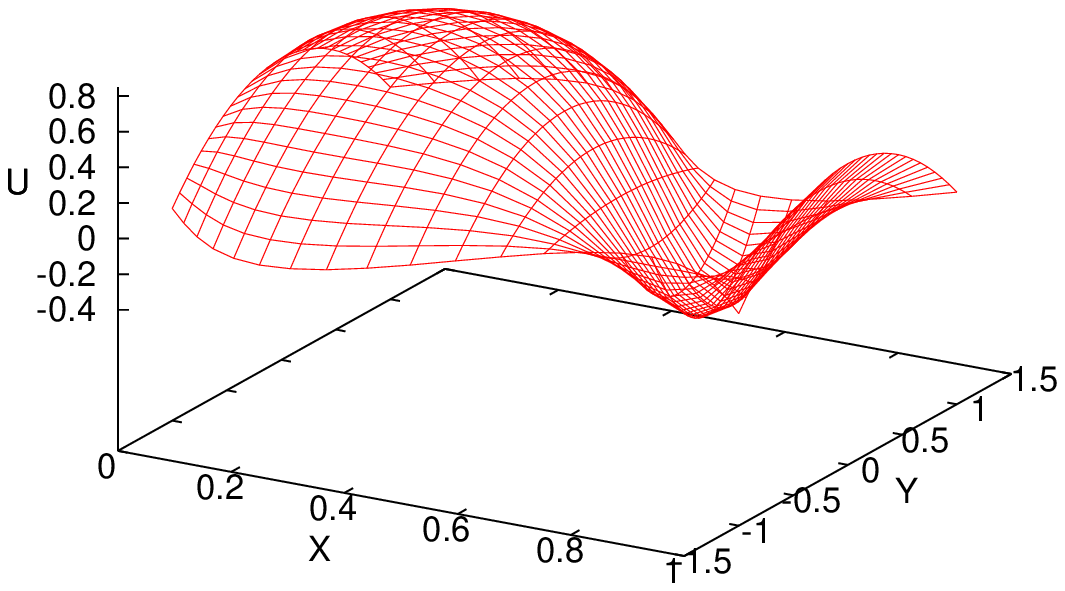}
\\$~~~~~~~~~~~~~~Fig.~3.1~(t=0.0)~~~~~~~~~~~~~~~~~~~~~~~~~~~~~~~~~~~~~~~~~~~~~~~~Fig.~3.2~(t=0.25)~~~~~~~~~~~~~~~~~~~~~~~~~~~~~~~~$\\
\includegraphics[width=3.3in,height=2.5in]{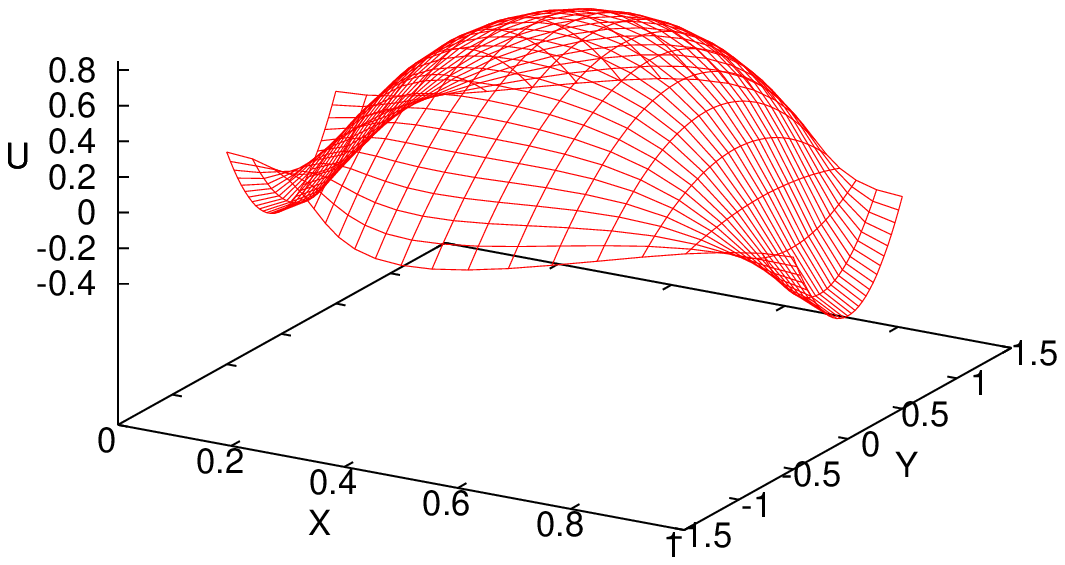}\includegraphics[width=3.3in,height=2.5in]{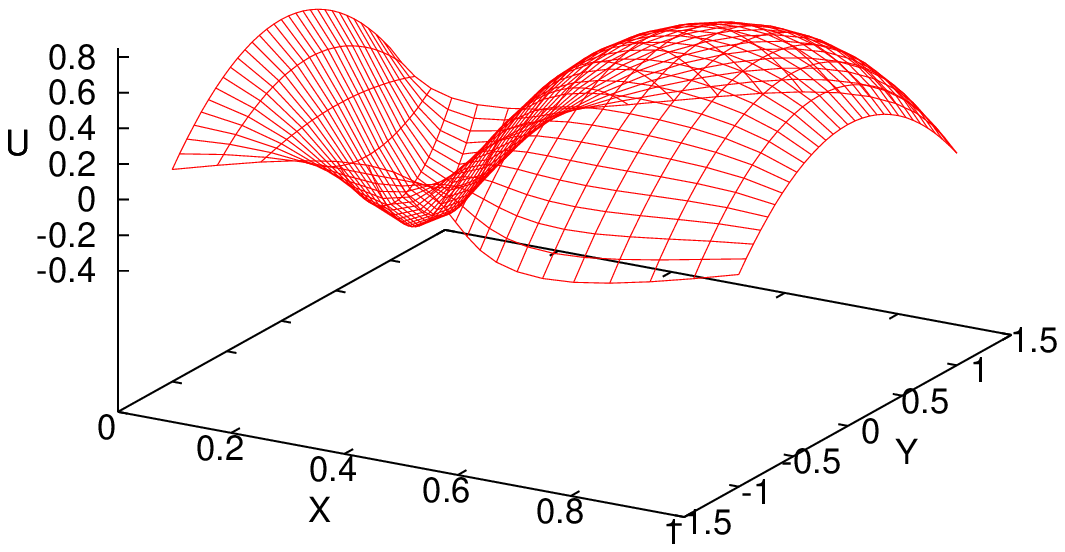}\\$~~~~~~~~~~~~~~ Fig.~3.3~(t=0.5)~~~~~~~~~~~~~~~~~~~~~~~~~~~~~~~~~~~~~~~~~~~~~Fig.~3.4~(t=0.75)~~~~~~~~~~~~~~~~~~~~~~~~~~~~~~~~~~~$\\
$~~~~~~~~~~~$Figs. 3.1-3.4: Aerial view of the velocity distribution at different times \\$~~~~~~~~~~~~~~~~~~~~~~~~~~~~~~~(\alpha=.003,~Da=10000,~\epsilon=1,~\Delta p=0,~\phi=.5$)
\end{figure}

\subsection{Pumping Characteristics}
Plots in Figs. 2 illustrate the variation of the volumetric flow rate
of peristaltic waves with pressure gradient for different values of
the amplitude ratio, couple stress parameter, permeability factor and
porosity parameter. Fig.2.1 shows that the results computed on the
basis of our study in the absence of permeability, porosity and couple
stress parameters perfectly match with the results reported by Shapiro
et al \cite{shapiro} for $\bar{Q}_{max}$ when $\Delta p$=0,
$\bar{Q}=Q/ac$. One may observe from Fig. 2.2 that in the range of
values of the pressure gradient examined in the present study, the
volumetric flow rate increases with the increase in the amplitude
ratio in the entire pumping region ($\Delta p >$0), in the free
pumping zone ($\Delta p =$0) as well as in the co-pumping region
($\Delta p <$0) for $\Delta p >-2.0$. However, the trend reverses as
soon as the pressure gradient drops below -2.0. These observations, of
course, hold good for a particular set of values indicated in
Fig. 2.2. Fig. 2.3 shows that in the entire pumping region the
volumetric flow rate increases with the increase in couple stress
parameter, whereas in the co-pumping region, a reverse trend is
noticed. From the same figure, we further derive the information that
for a Newtonian fluid ($\alpha\simeq0$), the magnitude of the pressure
rise is less than that for the couple stress fluid both in pumping and
co-pumping regions. One can observe from Fig. 2.4 that the volumetric
flow rate can be gradually reduced in the pumping region when $\Delta
p >5$ and enhanced in the pumping region when $\Delta p <5$, free
pumping region as well as in the co-pumping region by increasing the
Darcy number through some mechanism. It is found from Fig. 2.5 that
with a rise in the porosity parameter, volumetric flow rate increases in
the pumping region, but reduces in the co-pumping region.

\begin{figure}
 \includegraphics[width=2.2in,height=1.8in]{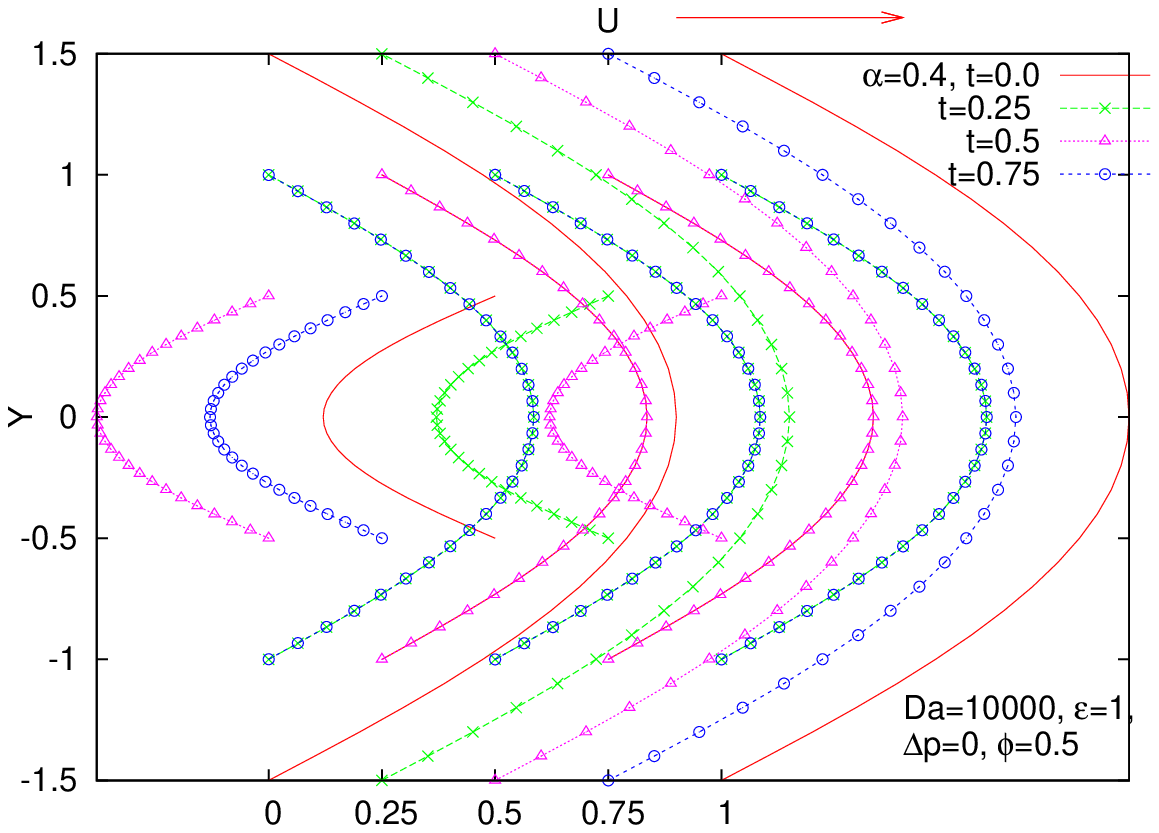}\includegraphics[width=2.2in,height=1.8in]{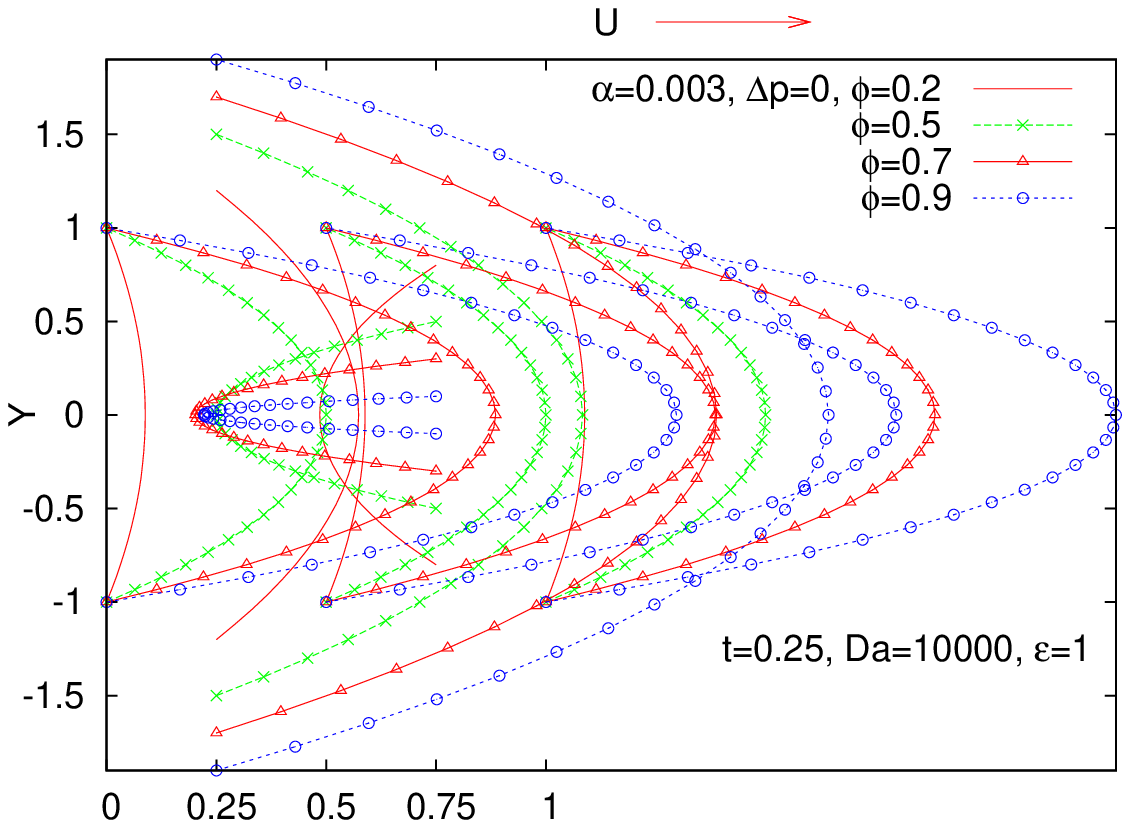}\includegraphics[width=2.2in,height=1.8in]{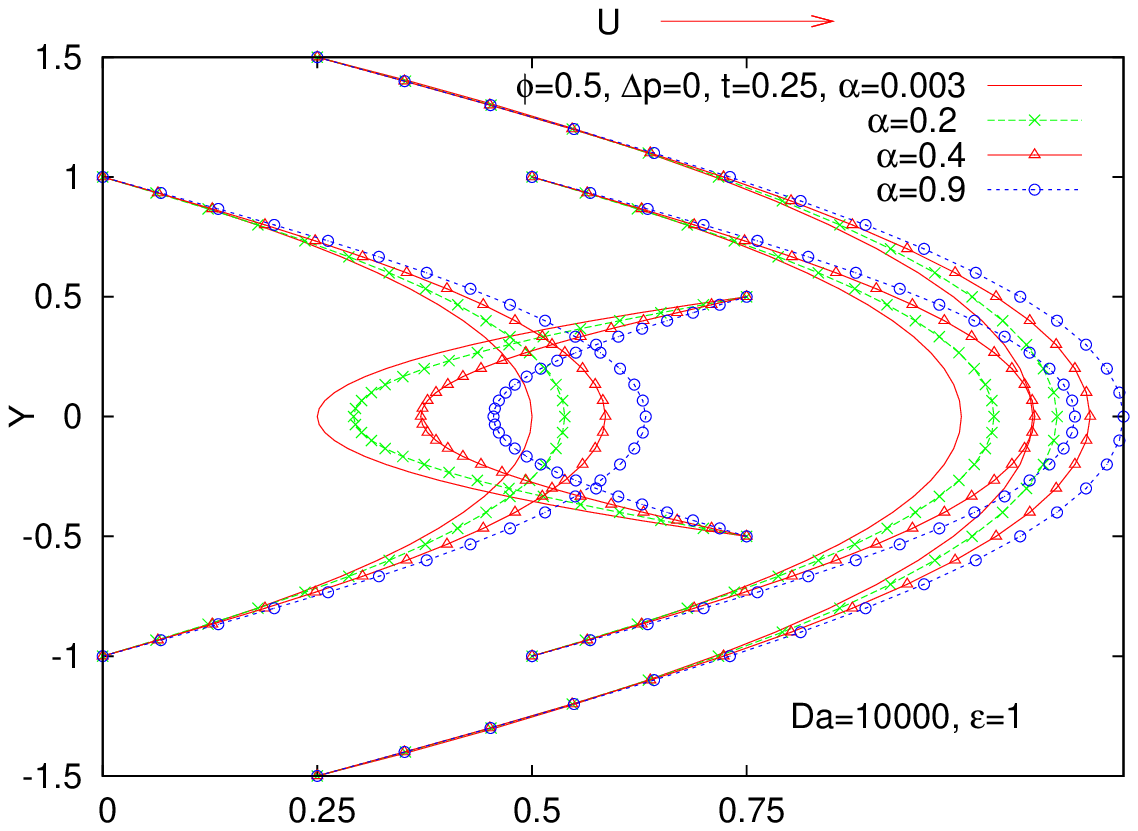}\\$~~~~~~~~~~~~~~ Fig.~3.5~~~~~~~~~~~~~~~~~~~~~~~~~~~~~~~~~~~~~~~~~~~~~~Fig.~3.6 ~~~~~~~~~~~~~~~~~~~~~~~Fig.~ 3.7~~~~~~~~~~$\\
\includegraphics[width=2.2in,height=1.8in]{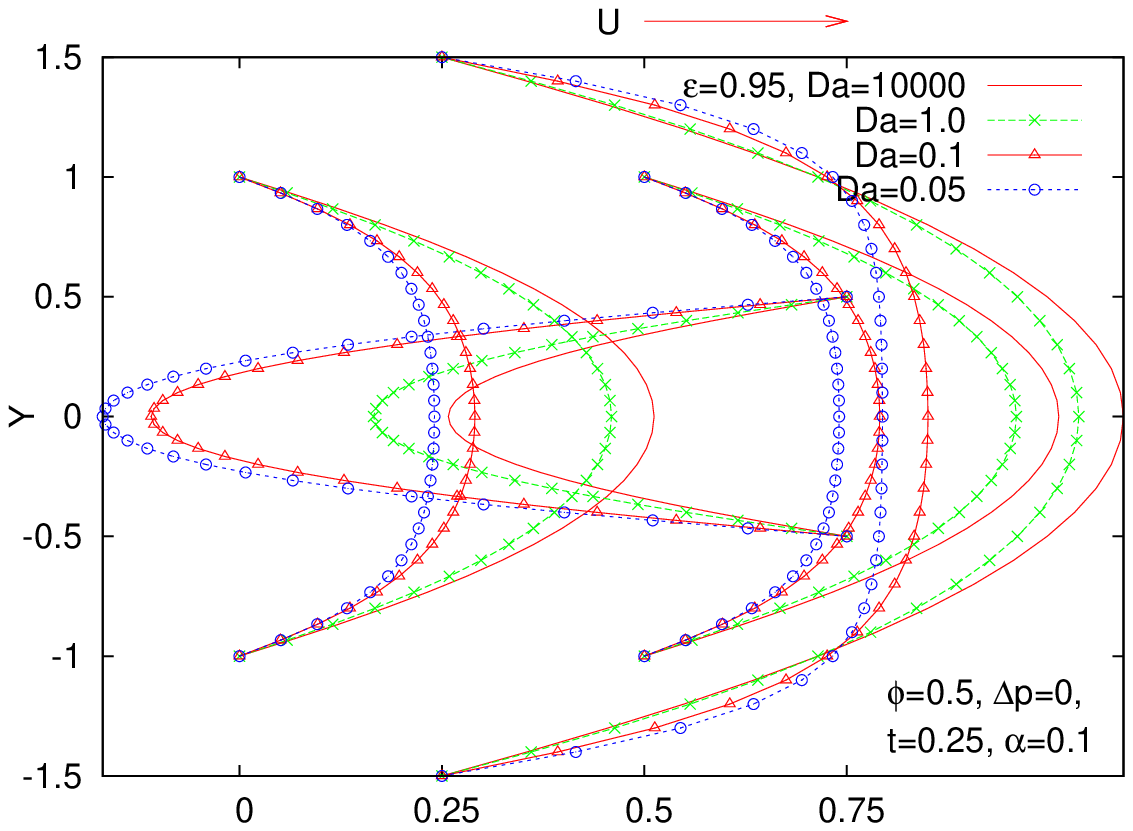}\includegraphics[width=2.2in,height=1.8in]{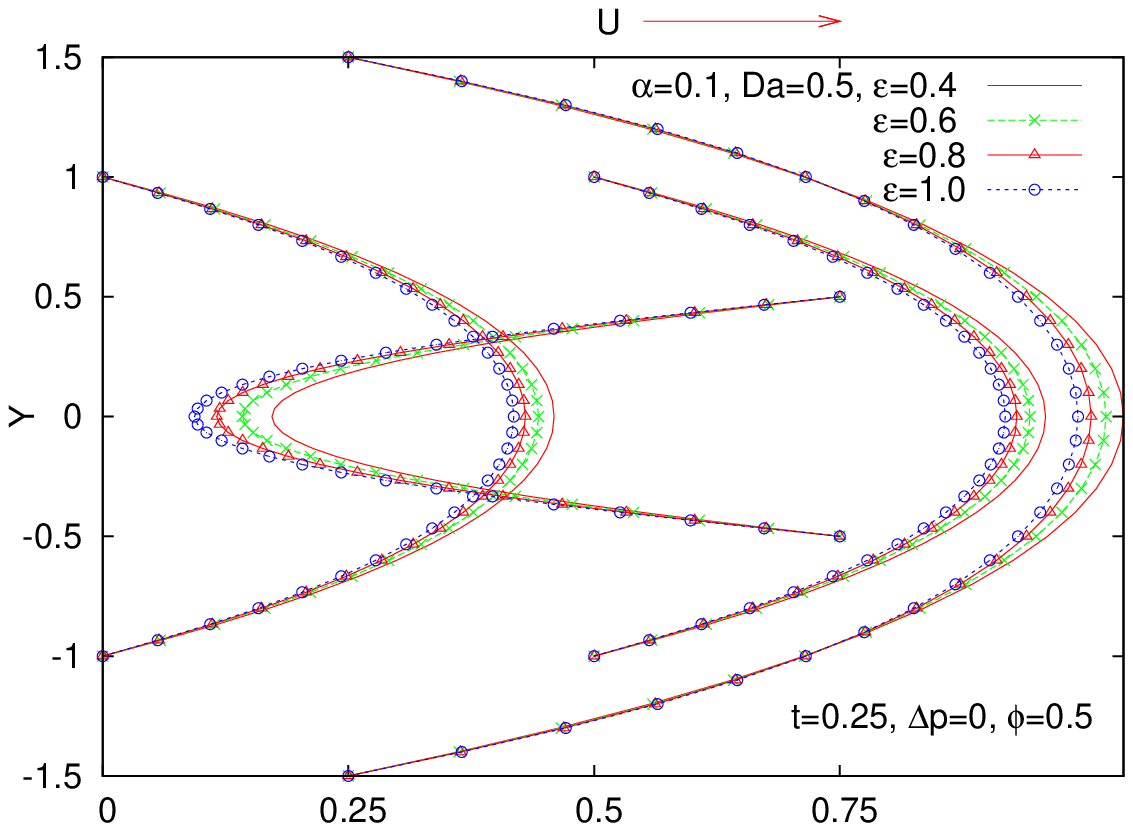}\includegraphics[width=2.2in,height=1.8in]{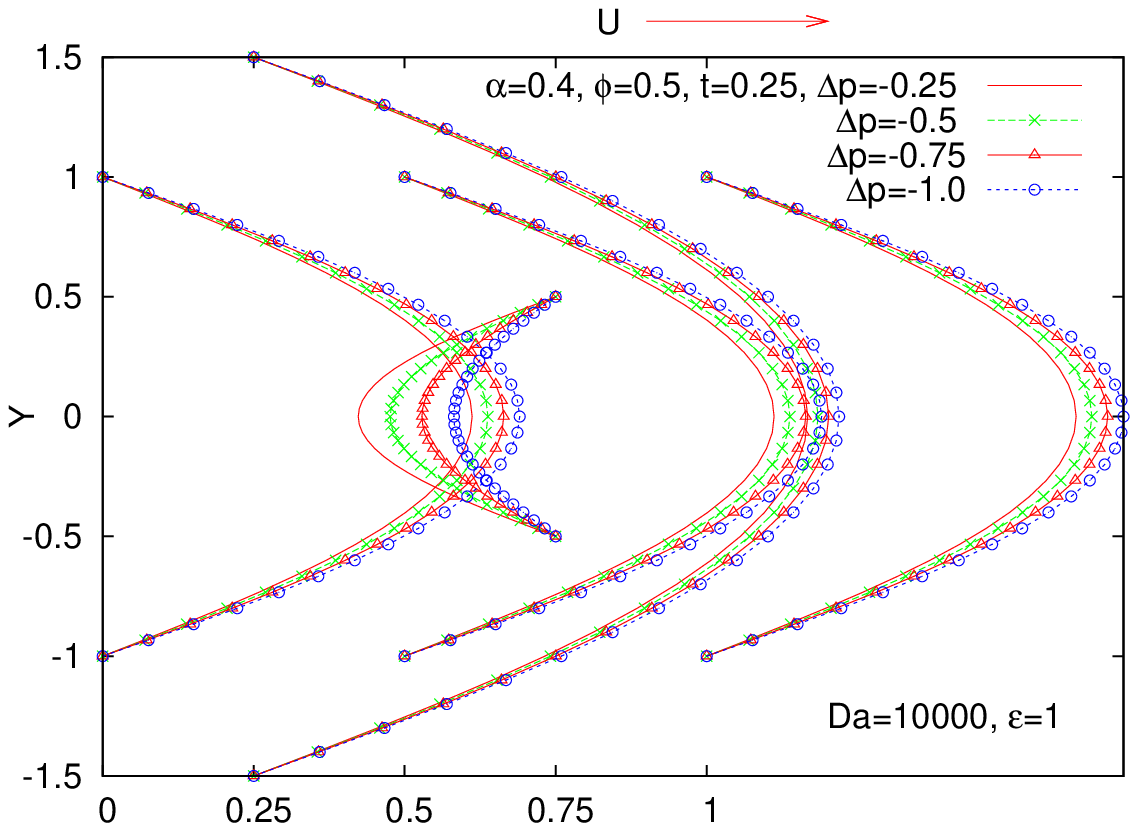}\\$~~~~~~~~~~~~~~ Fig.~3.8~~~~~~~~~~~~~~~~~~~~~~~~~~~~~~~~~~~~~~~~~~~~~~Fig.~3.9 ~~~~~~~~~~~~~~~~~~~~~~~Fig.~ 3.10~~~~~~~~~~$\\
\includegraphics[width=2.4in,height=1.8in]{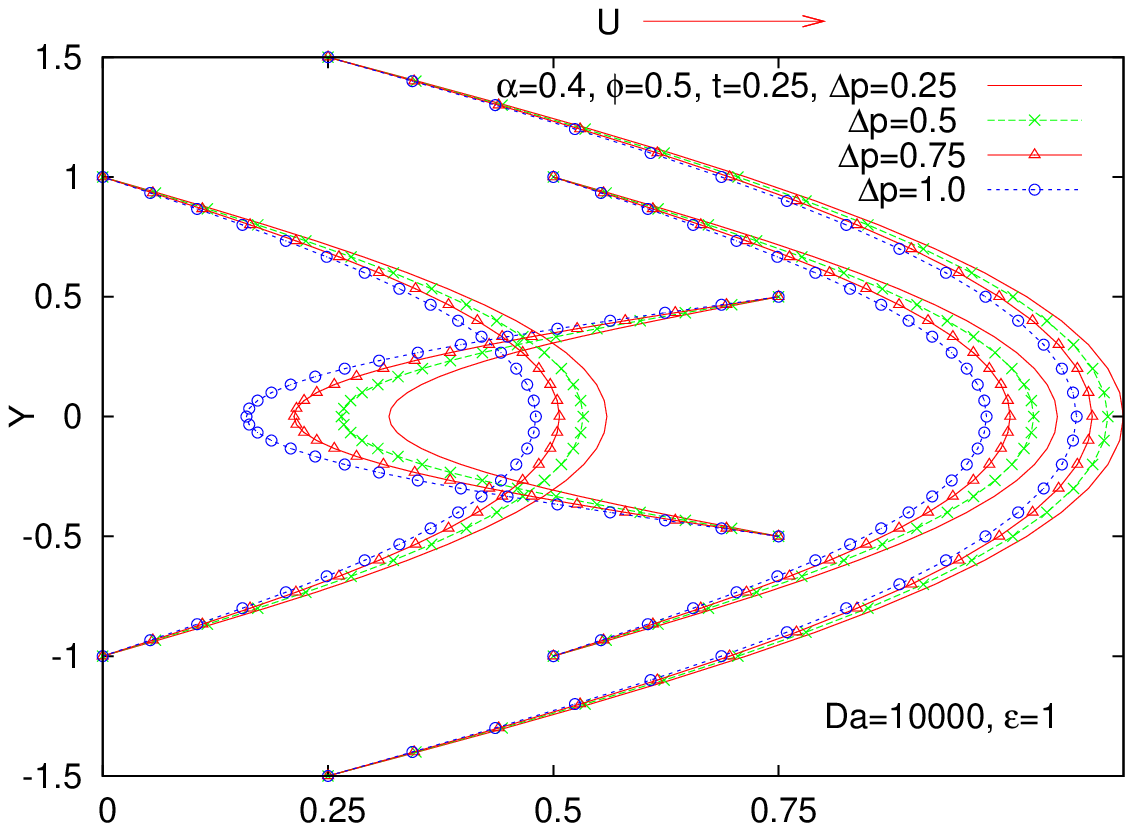}\includegraphics[width=2.4in,height=1.8in]{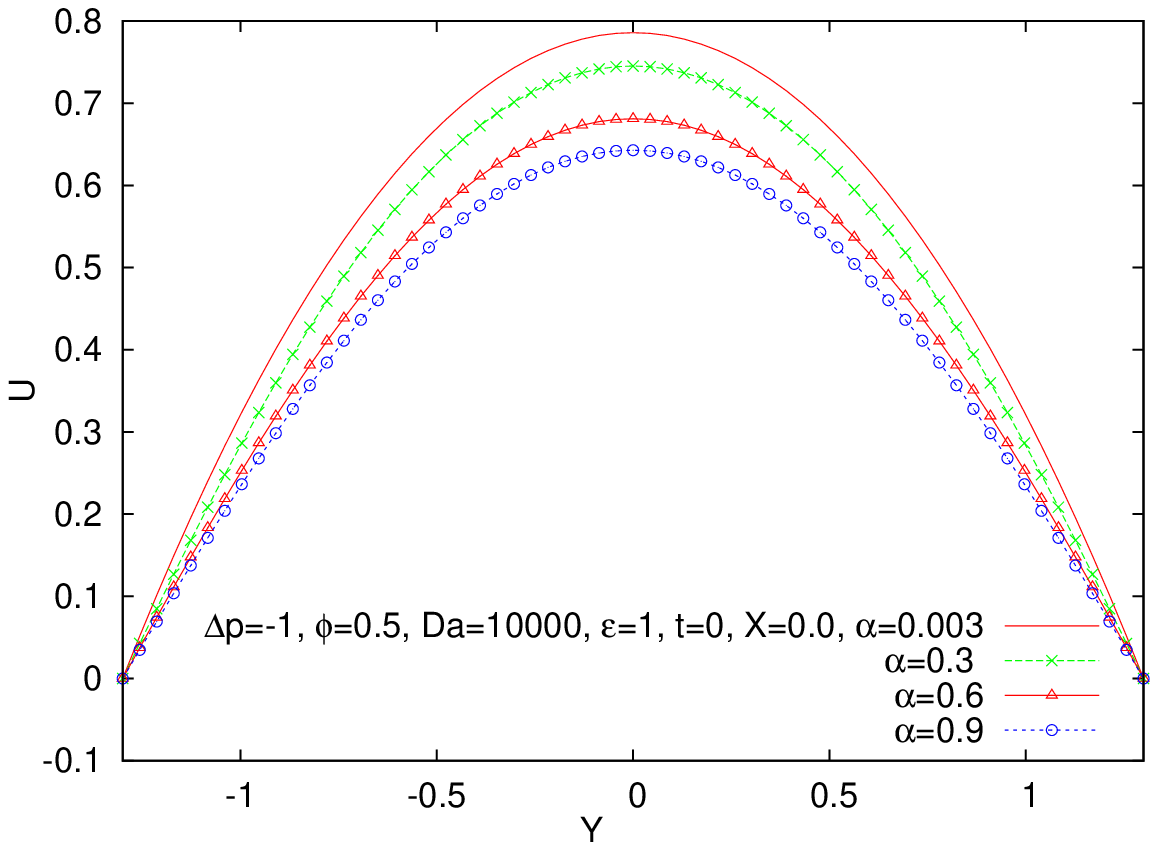}
$~~~~~~~~~~Fig.~3.11~~~~~~~~~~~~~~~~~~~~~~~~~~~~~~~~~~~~~~~~~~~~~~Fig.~3.12 ~~~~~~~~~~~~~~~~~~~~~~~~~~~~~~~$\\
$~~~~~~~~~~~$Figs. 3.5-3.12: Longitudinal velocity profiles in different situations
\end{figure}

 \subsection{Velocity Change}
An idea of the distribution of axial velocity can be from Figs 3 in
the cases of free pumping (Figs. 3.1-3.9), co-pumping (Fig. 3.10 and
3.12) and pumping (Fig. 3.11) zones for different values of the
amplitude ratio, the couple stress parameter, the Darcy number and the
porosity parameter. In the fixed frame of reference, since the
velocity profiles and the height of the channel change with time, the
axial velocity has been studied at the particular time instant
t=T/4. Aerial view of the velocity distribution of a Newtonian fluid
on the basis of our study has been presented in Figs 3.1-3.4, while
Fig. 3.5 gives the velocity distribution of a couple stress fluid
($\alpha=0.4$) in the plane of the channel. We observe that at any
instant of time there exists a retrograde flow region. But the forward
flow region is predominant here as the time averaged flow rate
$\bar{Q}$ is positive. Our study also reveals that for a couple stress
fluid, there exist two stagnation points; for example, at t=0, one of
them lies between X=0.25 and X=0.5, and the other between X=0.5 and
X=0.75. A similar observation was reported earlier for a similar study
\cite{takabatake1}. Figs 3.6-3.7 reveal that for free pumping case
($\Delta p=0$), the magnitude of the velocity for Newtonian as well as
couple stress fluid in the forward and retrograde flow regions
increases when the value of the amplitude ratio increases. Moreover,
from Fig 3.7, one may observe that as the value of $\alpha$ increases,
the velocity in the forward flow region increases. It may also be
noted that the couple stress parameter $\alpha$ enhances the flow
velocity in the retrograde flow region. The influence of the Darcy
number Da on the velocity profile for a couple stress fluid (with
$\alpha=0.1$) is depicted in Fig. 3.8.  One can further note that as
the Darcy number Da decreases, the axial velocity diminishes in
forward region, whereas it increases in magnitude in retrograde flow
region (i.e  retrograde flow enhances). It is worthwhile to mention here
that the parabolic nature of the velocity profile is disturbed in the
central region for small values of Da. We further observe that for a
couple stress fluid, the axial velocity decreases with an increase in
the value of the porosity parameter $\epsilon$ of the channel
(cf. Fig3.9) in both forward and backward flow regions.
Figs. 3.10-3.11 give the distribution of velocity of a couple stress
fluid for co-pumping ($\Delta p<0$) and pumping ($\Delta p>0$)
cases. From both the figures, it is revealed that as the value of
$\Delta p$ decreases, the velocity increases in both the flow
regions. It is important to mention that unlike free pumping case, in
the co-pumping case with $\Delta p=-1$, the velocity reduces with the
increase in the value of the couple stress parameter $\alpha$
(cf. Fig. 3.12).

\begin{figure}
\begin{center}
 \fbox{\includegraphics[width=3.2in,height=2.0in]{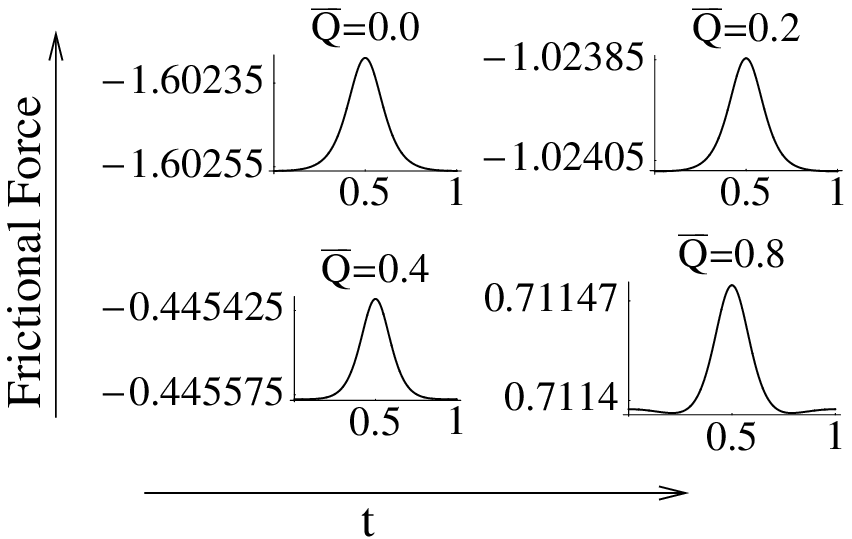}}\fbox{\includegraphics[width=3.2in,height=2.0in]{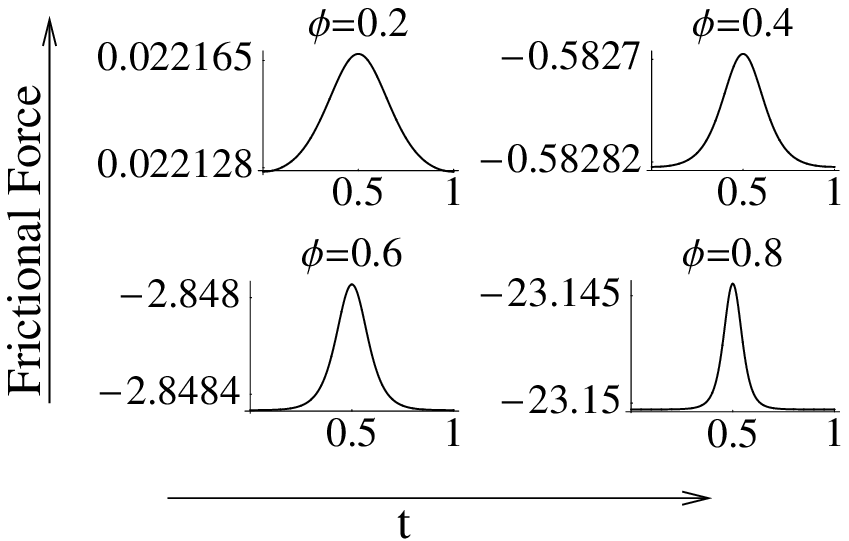}}\\$~~~~~~~~~~~~~~(a)~~~~~~~~~~~~~~~~~~~~~~~~~~~~~~~~~~~~~~~~~~~~~~~~~~~~(b)~~~~~~~~~~~~~~~~~~~~~~~~~~~~~~~~$\\ \fbox{\includegraphics[width=3.2in,height=2.0in]{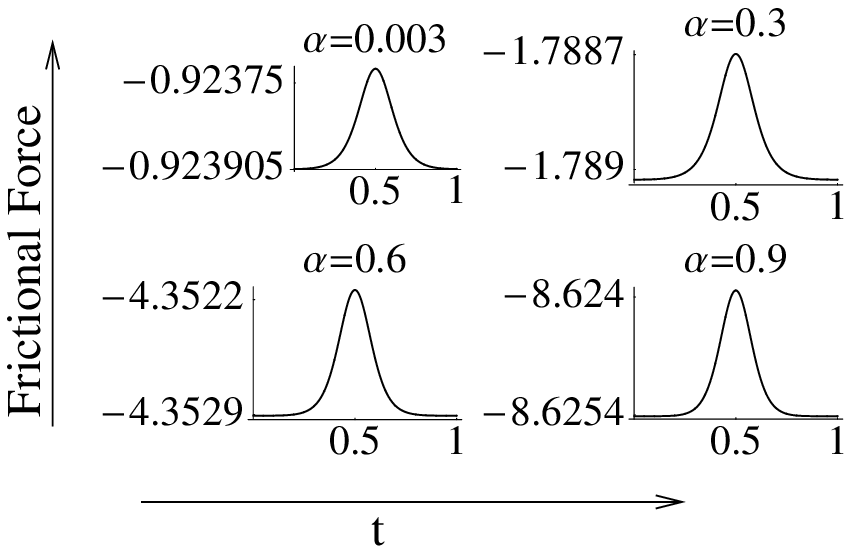}}\fbox{\includegraphics[width=3.2in,height=2.0in]{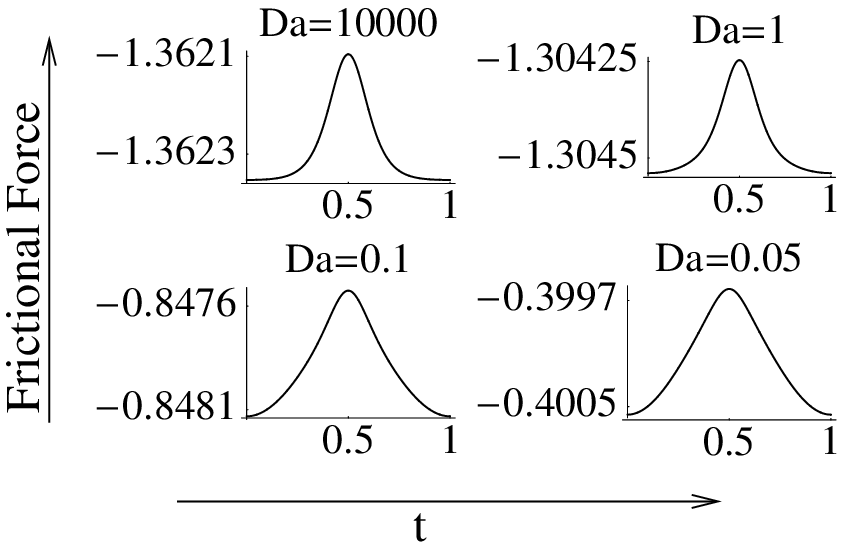}}\\$~~~~~~~~~~~~~~(c)~~~~~~~~~~~~~~~~~~~~~~~~~~~~~~~~~~~~~~~~~~~~~~~~~~~~(d)~~~~~~~~~~~~~~~~~~~~~~~~~~~~~~~~$\\
\fbox{\includegraphics[width=3.6in,height=2.0in]{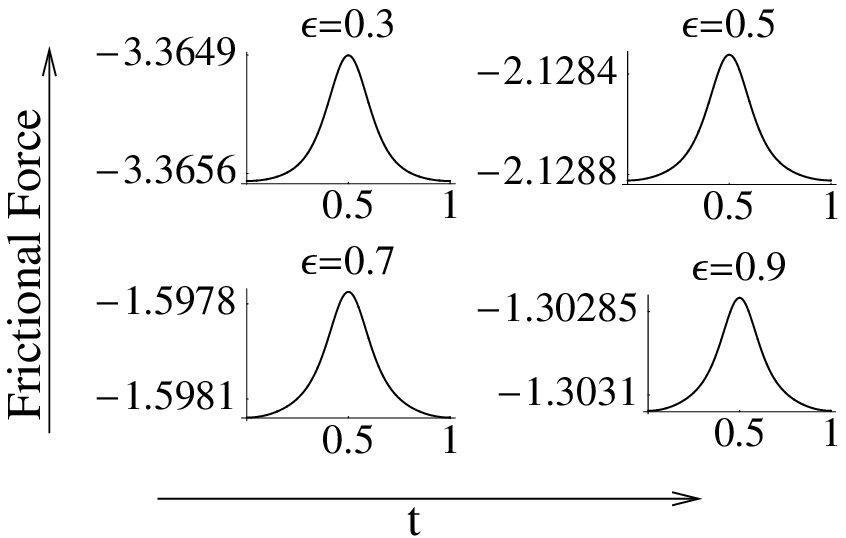}}\\(e)\\ Figs. 4:
 Frictional force versus time for different cases\\ (a)
 $\alpha=0.2,~\phi=0.5,~Da=10000,~\epsilon=1$\\ (b)
 $\alpha=0.2,~\bar{Q}=0.1,~Da=10000,~\epsilon=1$ \\(c)
 $\phi=0.5,~\bar{Q}=0.1,~Da=10000,~\epsilon=1$ \\(d)
 $\alpha=0.2,~\phi=0.5,~\bar{Q}=0.1,~\epsilon=0.95$\\ (e)
 $\alpha=0.2,~\phi=0.5,~\bar{Q}=0.1,~Da=0.5$
\end{center}
 \end{figure}

 \subsection{Frictional Force}
The average rise in frictional force in the case of a couple stress
fluid are calculated over one wave period. Figs. 4.(a) depict the
variation of frictional force with time for different values of
$\bar{Q}~when~\phi=0.5$. It can be observed that the effect of increasing
the flow rate $\bar{Q}$ is to enhance the frictional force F. As
expected, from Figs. 4(b), we find that F decreases with rise in amplitude
ratio. Figs. 4(c) reveal that the magnitude of the frictional force
for a Newtonian fluid is less than that in the case of a couple stress
fluid and that as $\alpha$ increases, the magnitude of F enhances. The
quantum of influence of the Darcy number on frictional force is shown
in Figs. 4(d). These figures indicate that the magnitude of F
diminishes when the value of Da reduces, while Figs. 4(e) show that with
an increase in the porosity parameter $\epsilon$, the magnitude of the
frictional force decreases.
\begin{figure}
\includegraphics[width=3.5in,height=2.4in]{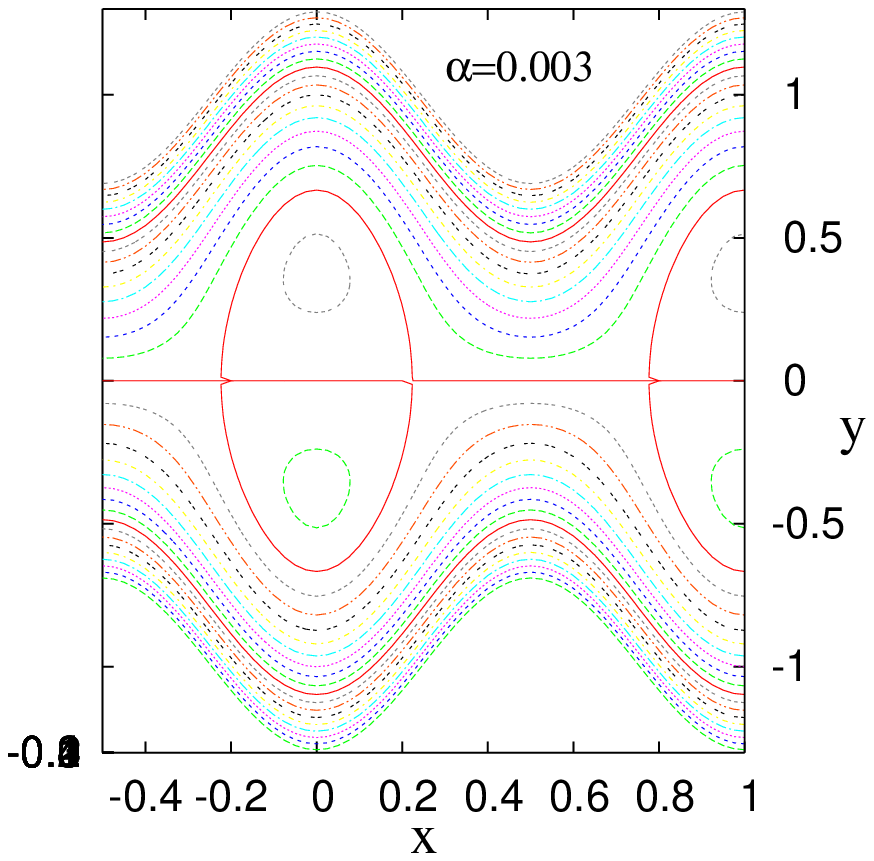}\includegraphics[width=3.5in,height=2.4in]{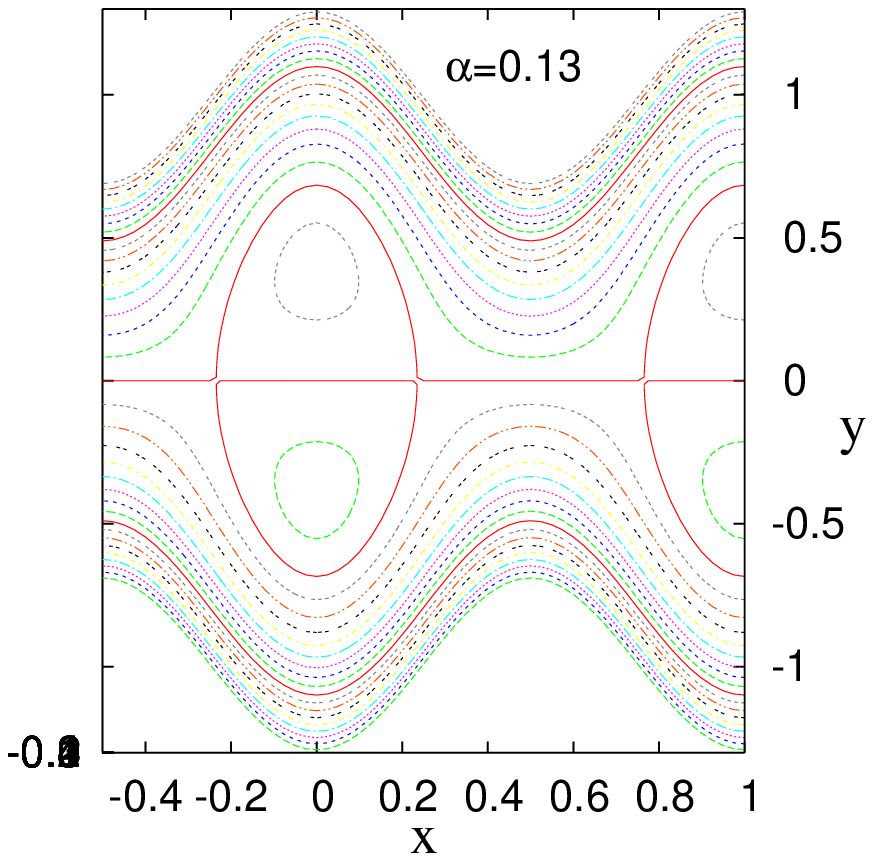}\\
\includegraphics[width=3.5in,height=2.4in]{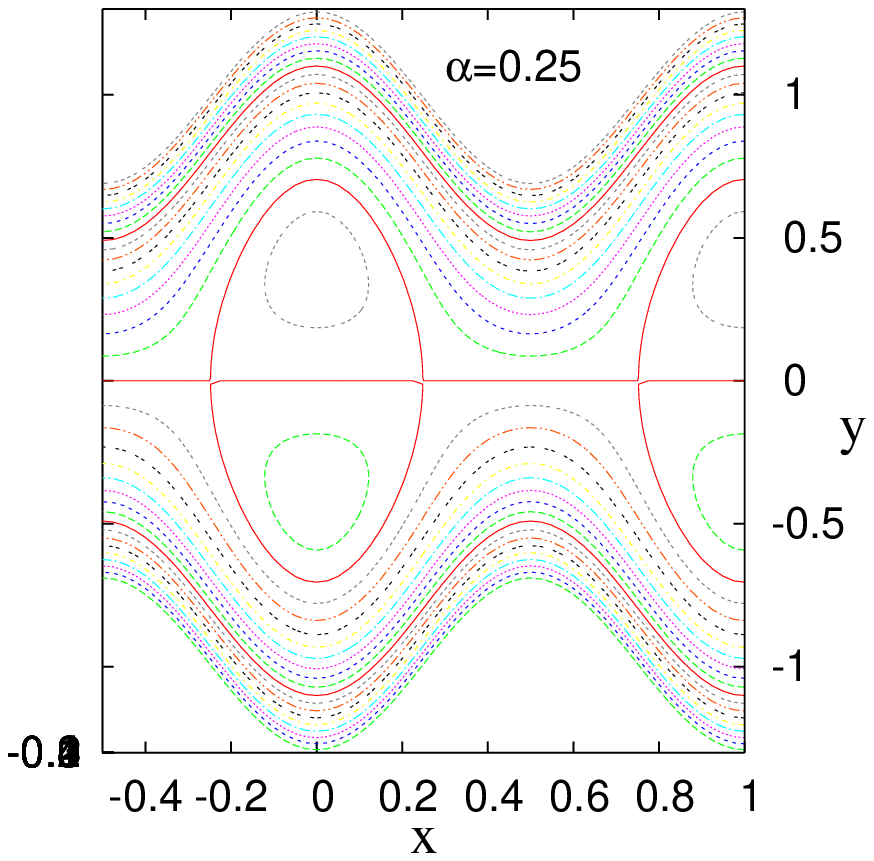}\includegraphics[width=3.5in,height=2.4in]{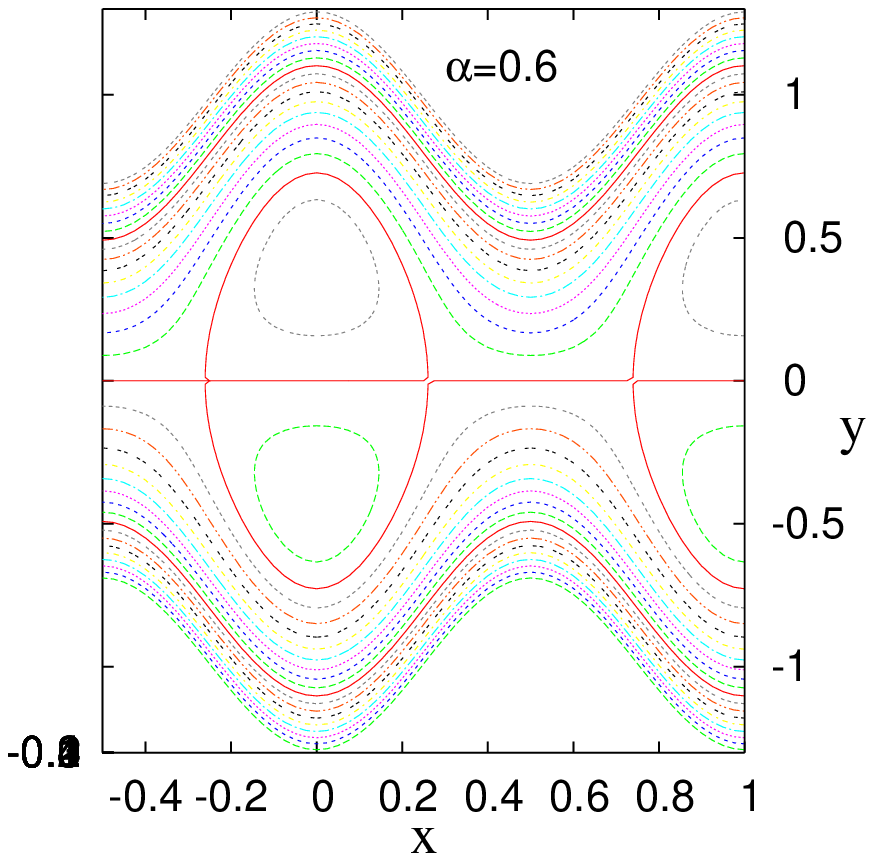}\\
Figs 5.1: Streamline patterns for different values of $\alpha~ with~ \bar{Q}=1.3,~ \phi=0.3,~ Da=10000,~ \epsilon=1$
\end{figure}

\subsection{Streamline Patterns and Trapping}
Under certain conditions the streamlines on the center line in the
wave frame of reference are found to split in order to enclose a bolus
of fluid particles circulating along closed streamlines. This
phenomenon is referred to as trapping, which is a characteristic of
peristaltic motion. Since this bolus appears to be trapped by the
wave, the bolus moves with the same speed as that of the
wave. Figs. 5.1 illustrate the streamline patterns and trapping for
different values of the couple stress parameter $\alpha$ corresponding
to $\bar{Q}=1.3,~ \phi=0.3,~ Da=10000,~\epsilon=1$. It is important to
observe that the size of the trapped bolus increases as the couple
stress parameter $\alpha$ increases. Formation of trapping zone and
streamline patterns is depicted in Figs. 5.2 for different values of
Darcy number Da, when
$\alpha=0.003,~\epsilon=0.8,~\phi=0.5,~\bar{Q}=1.4$. These figures
reveal that with a reduction in Da, the trapping zone decreases and it
disappears completely when Da attains the value 0.08. Plots showing
the effect of the porosity parameter $\epsilon$ on trapping are
presented in Figs. 5.3. One can observe that size of the trapped bolus
decreases with increase in $\epsilon$ and vanish when
$\epsilon$=1. Figs. 5.4 indicate that size of trapped bolus increases
as the flow rate increases and that with increasing flow rate, the
bolus gets shifted towards the boundary more and more.
\begin{figure}
\includegraphics[width=3.5in,height=2.4in]{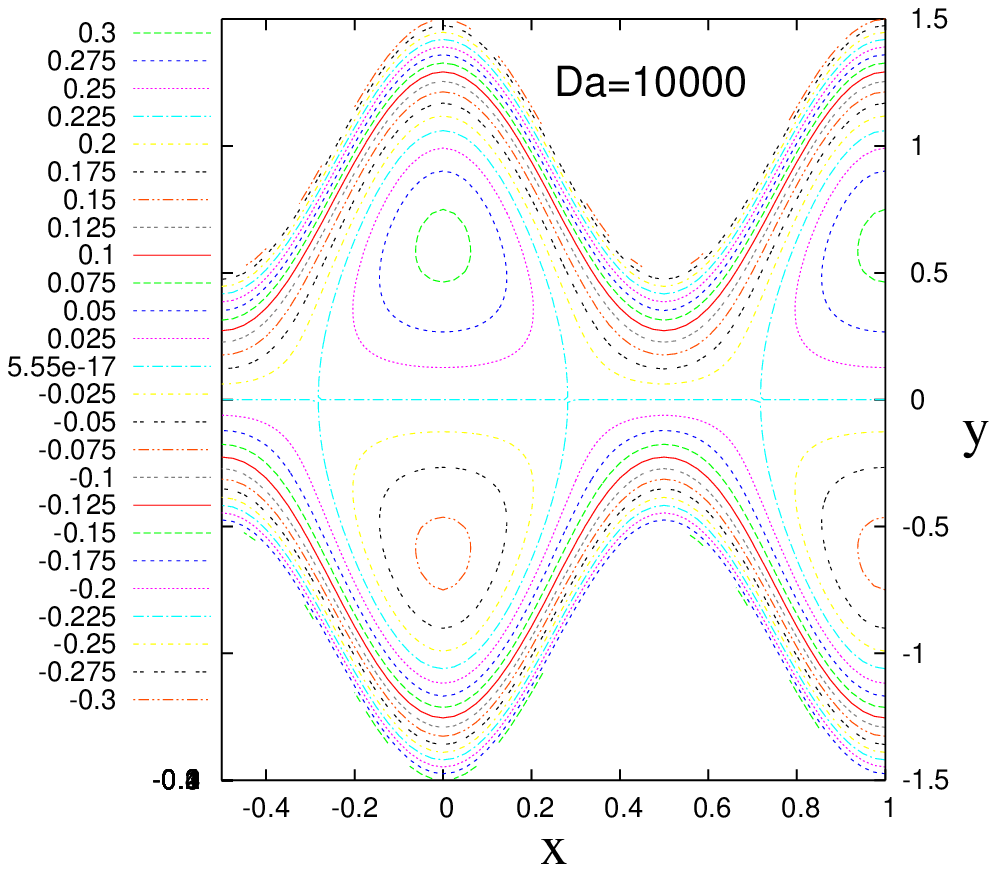}\includegraphics[width=3.6in,height=2.4in]{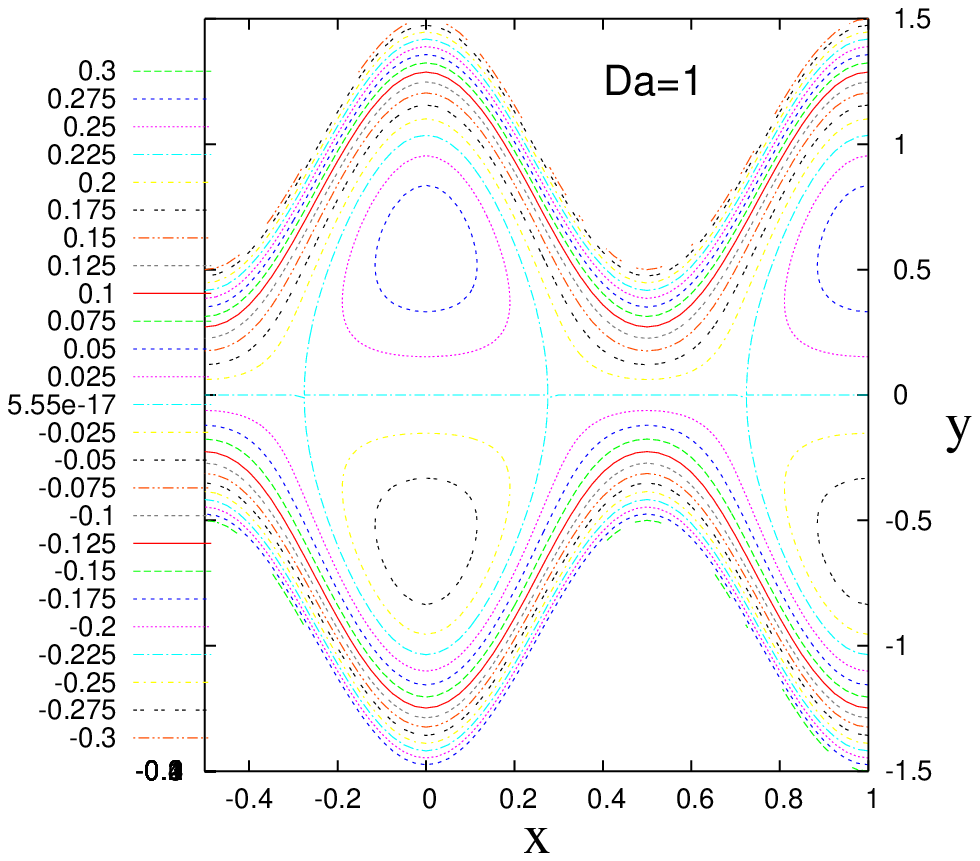}\\
\includegraphics[width=3.6in,height=2.4in]{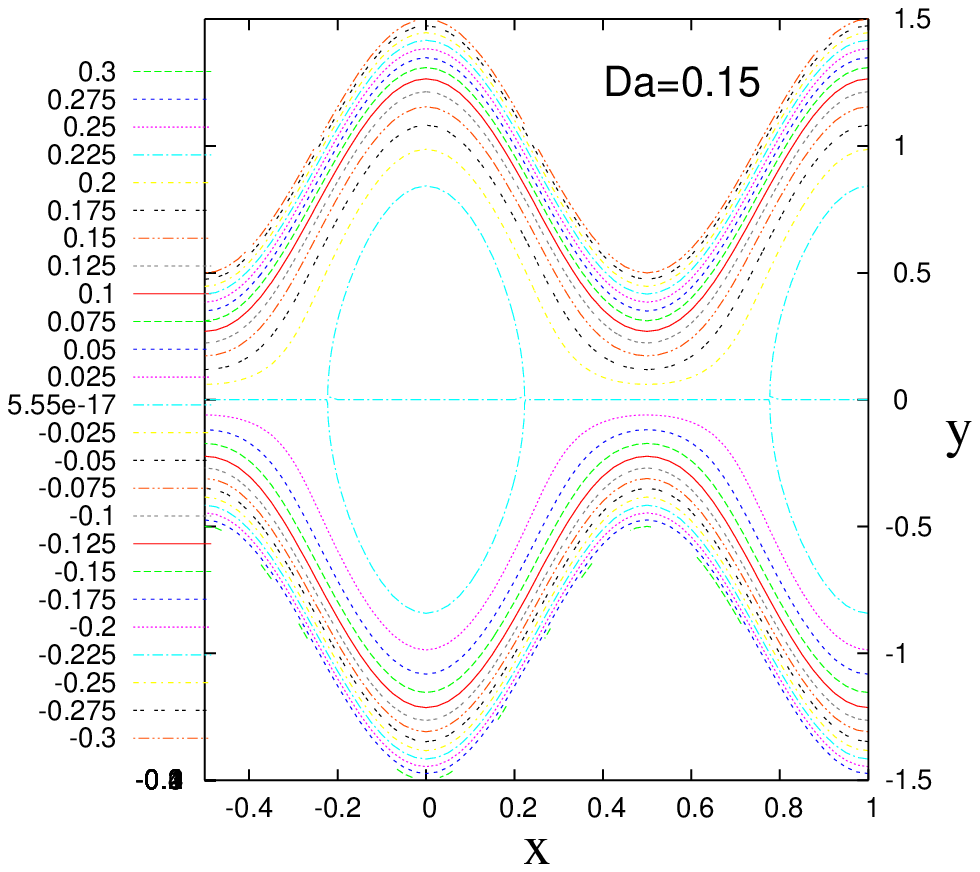}\includegraphics[width=3.6in,height=2.4in]{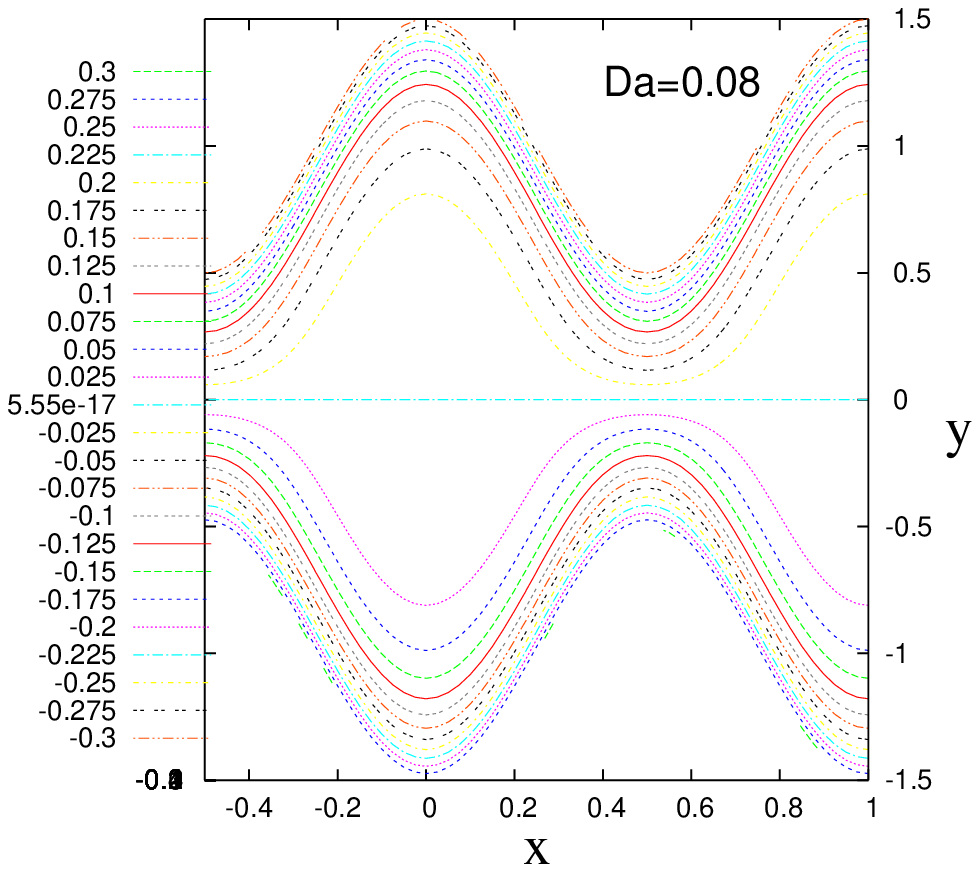}\\
Figs. 5.2: Streamline patterns for different values of Da with $\alpha=0.003,~ \bar{Q}=1.4,~ \phi=0.5,~ \epsilon=0.8$
\end{figure}
\begin{figure}
\includegraphics[width=3.5in,height=2.4in]{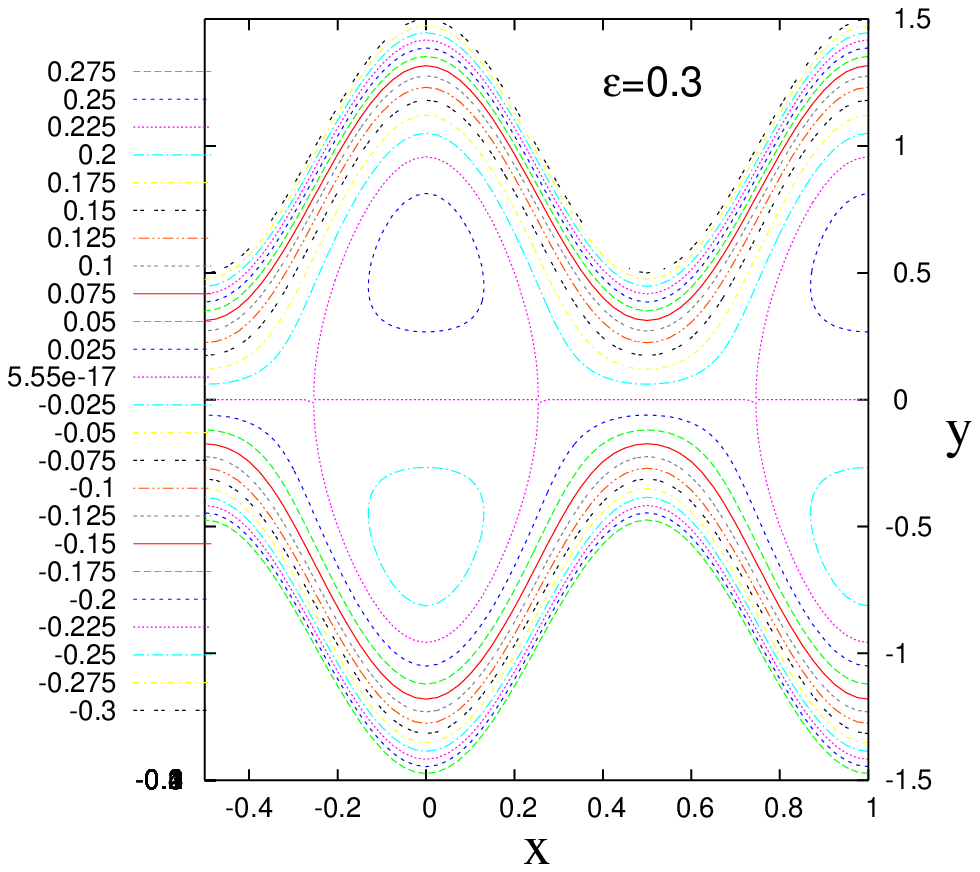}\includegraphics[width=3.5in,height=2.4in]{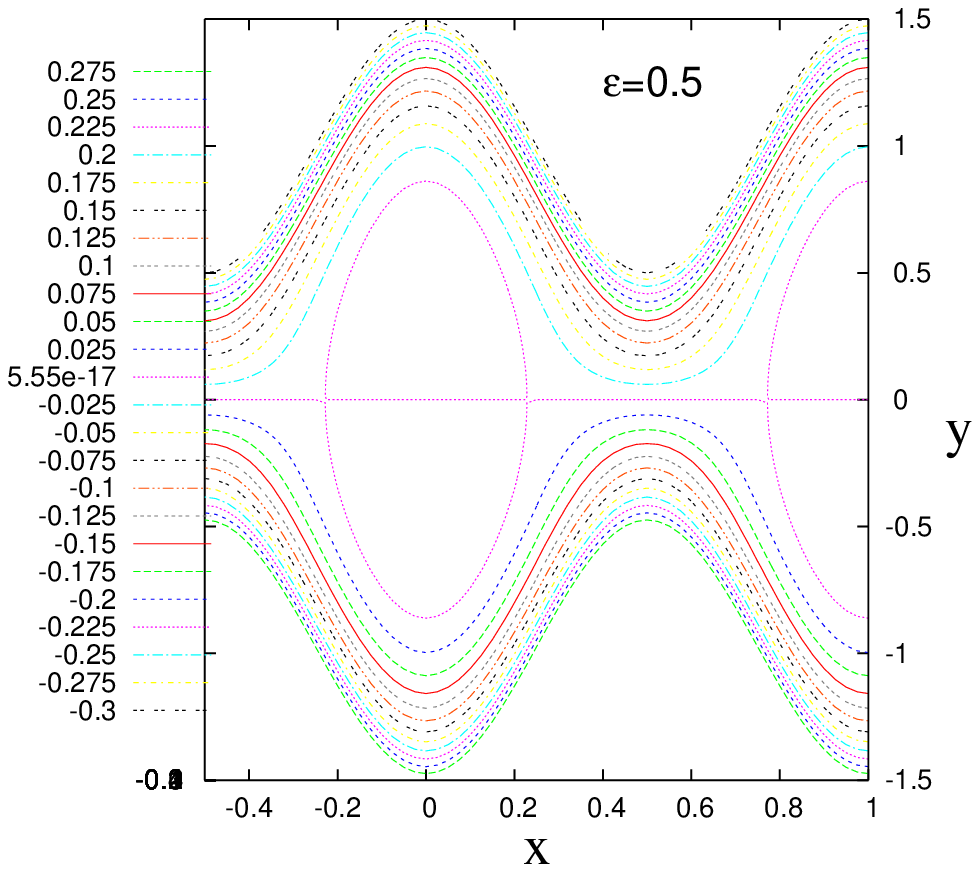}\\
\includegraphics[width=3.5in,height=2.4in]{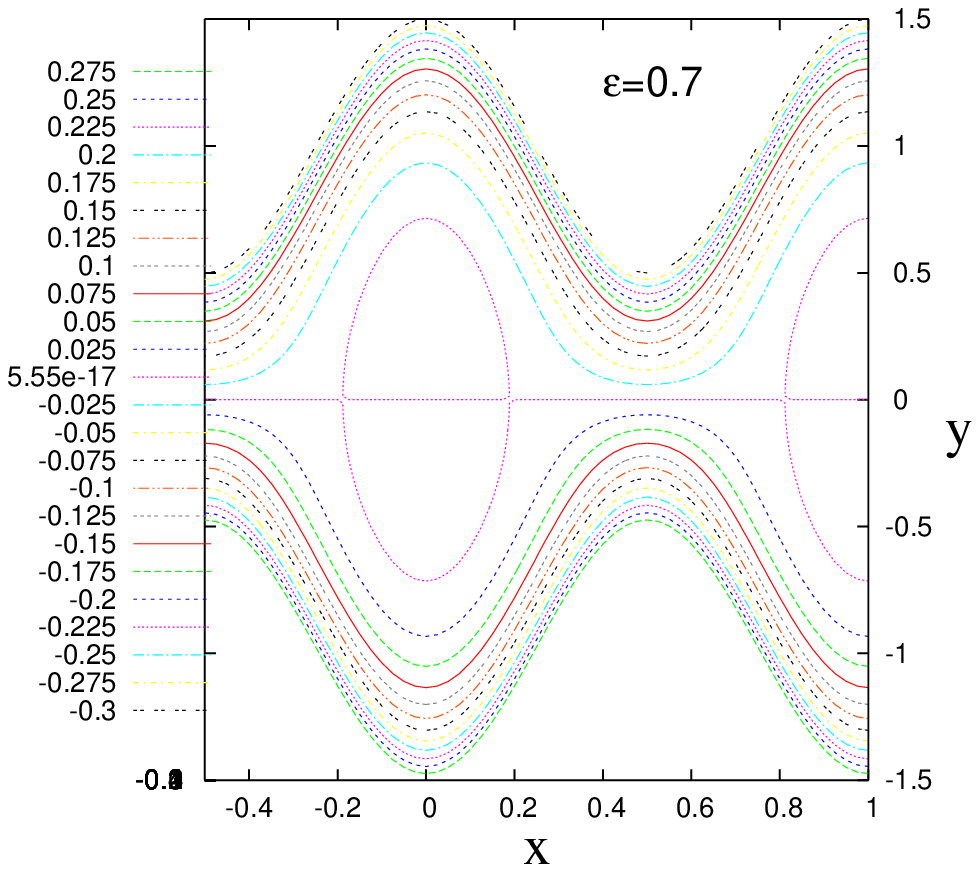}\includegraphics[width=3.5in,height=2.4in]{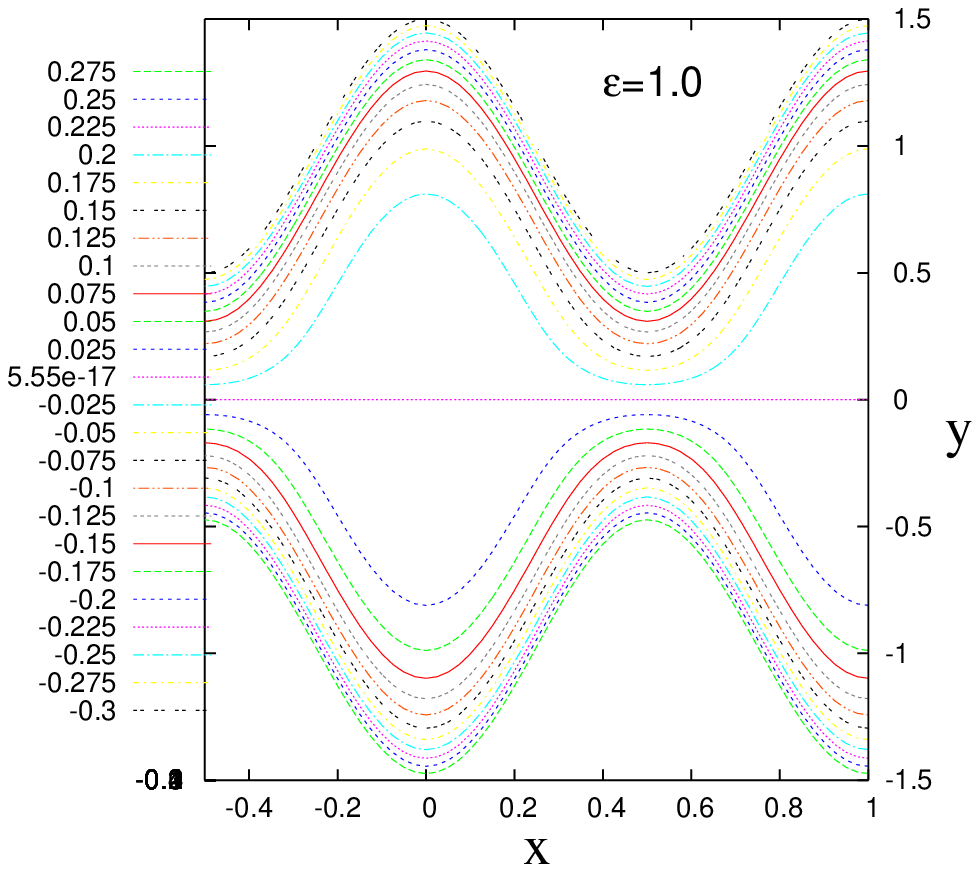}\\
Figs. 5.3: Streamline patterns for different values of $\epsilon~ with~ \alpha=0.003,~ \bar{Q}=1.4,~ \phi=0.5,~ Da=0.1$
\end{figure}
\begin{figure}
\includegraphics[width=3.5in,height=2.4in]{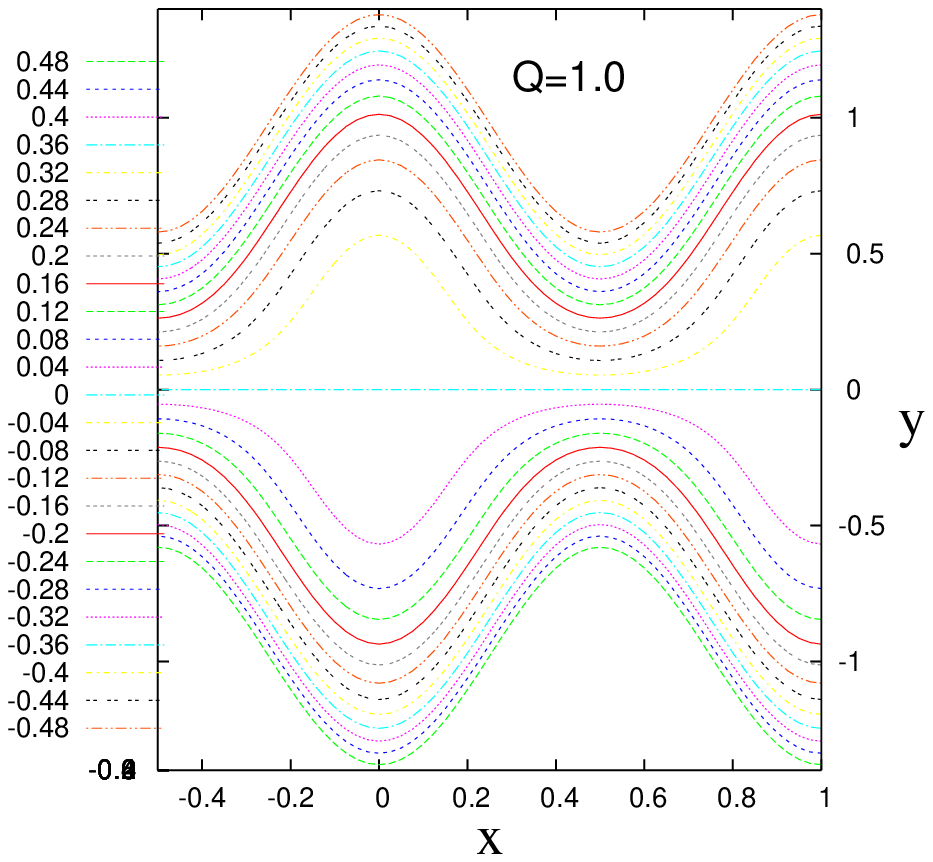}\includegraphics[width=3.5in,height=2.4in]{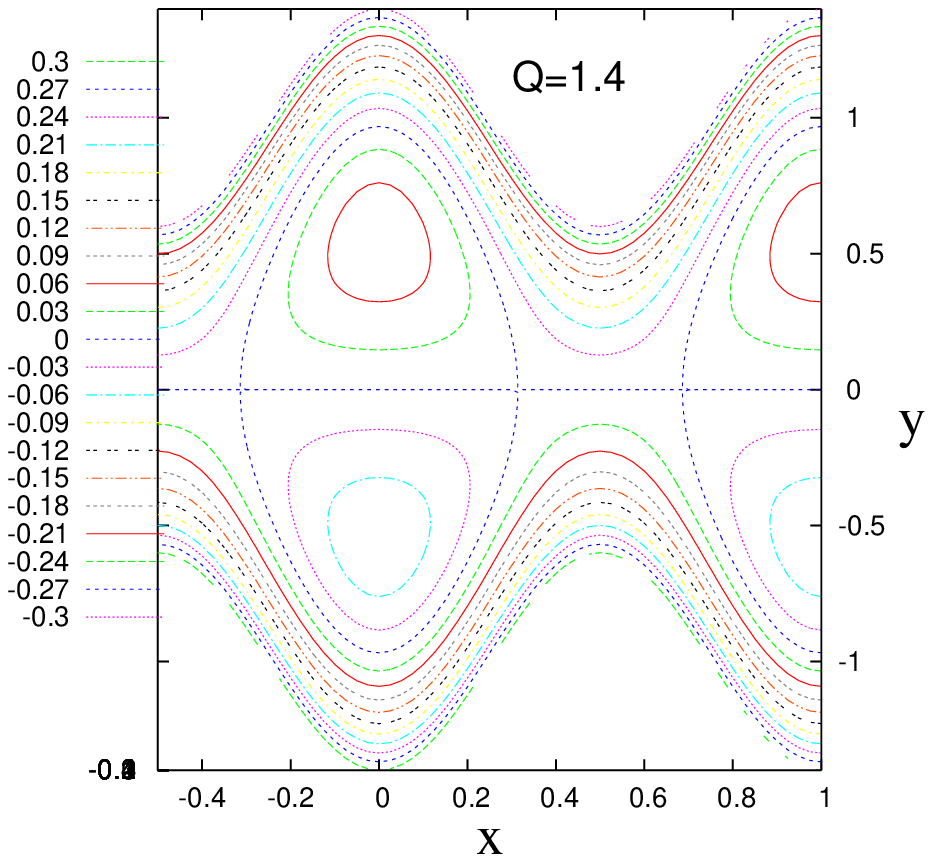}\\
\includegraphics[width=3.5in,height=2.4in]{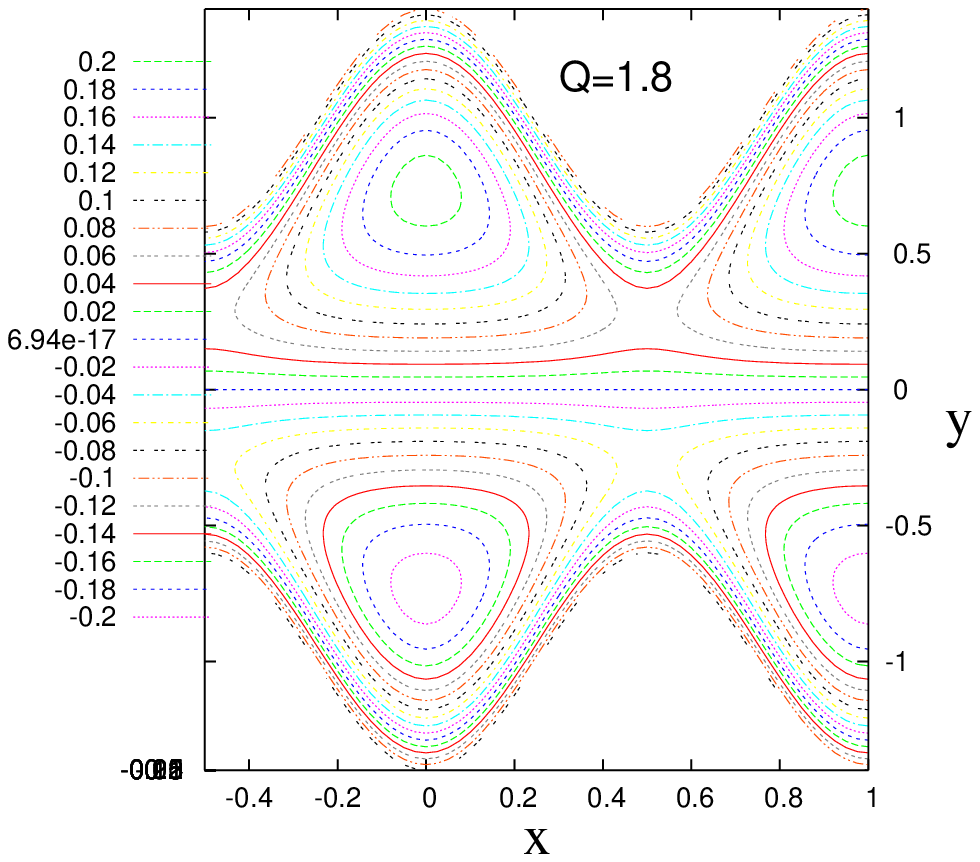}\includegraphics[width=3.5in,height=2.4in]{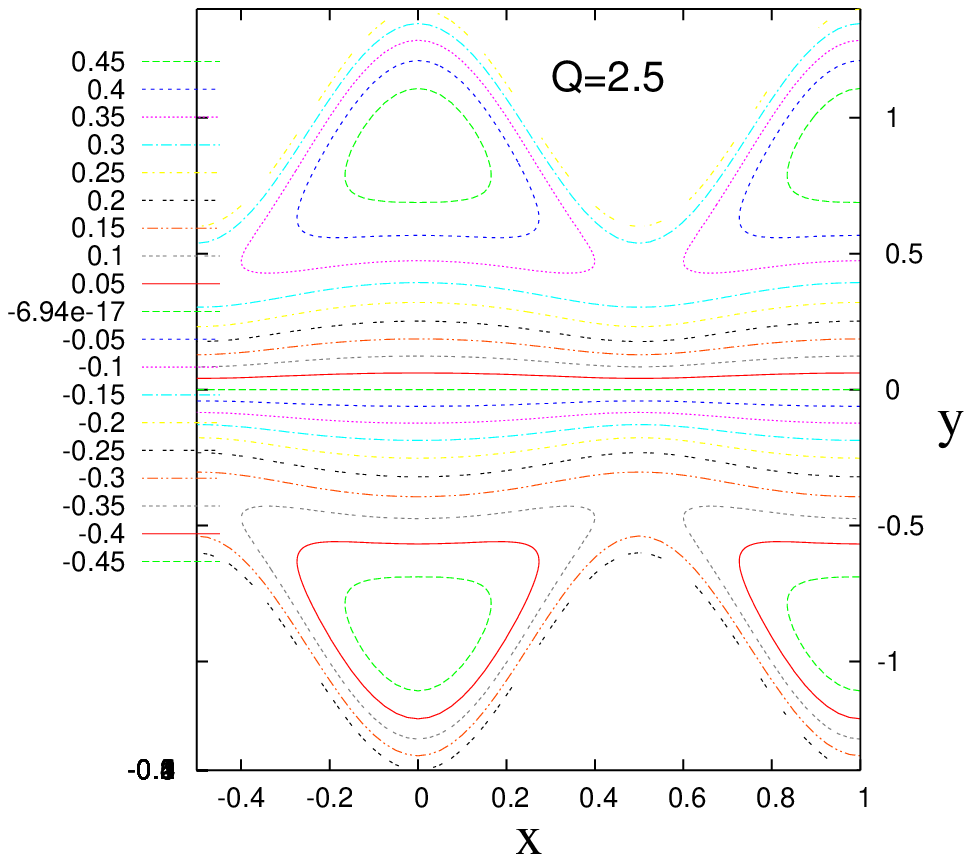}\\
Figs. 5.4: Streamline patterns for different values of $\bar{Q}~ with~ \alpha=0.003,~ \phi=0.4,~ Da=10000,~ \epsilon=1$
\end{figure}

\section{Summary and Conclusion}
Peristaltic transport of a couple stress fluid in a porous medium has
been the concern of the present investigation. The study particularly
pertains to a situation when the Reynolds number is low and curvature
of the channel is quite small. The primary motivation behind the study
has been to have an insight into the physiological problem of
peristaltic transport of blood in a segment of the micro-circulatory
system, when the segment is in a pathological state, whereby the lumen
of the particular segment turns into a porous structure. Emphasis has
been paid to investigate the peristaltic pumping characteristics,
blood velocity, frictional force, streamline patterns and trapping as
a function of the peristaltic wave amplitude as well as Darcy
permeability and porosity parameters. Although the analysis starts
with an analytical procedure, suitable numerical techniques have been
employed whenever analytical methods have been found inadequate. The
results and observations of this study (reduced to some particular
cases) are in good agreement with those reported earlier by other
investigators (cf. \cite{shapiro}). A very significant conclusion that
can be made on the basis of this study is that it is possible to
enhance the pressure rise $\Delta p$ as well as the peristaltic
pumping performance by increasing the amplitude ratio/couple stress
parameter and/or by reducing the porosity. The study further reveals
the important fact that change in velocity is strongly affected by the
wave amplitude, the couple stress effect and the porosity factor. The
other important conclusion that can be drawn out of this study is that
while flow reversal is enhanced due to increase in pressure rise in
the free pumping case, it is possible to bring about reduction in flow
reversal by increasing the couple stress
effect/permeability. Moreover, the occurrence of trapping can be
reduced/eliminated by decreasing the permeability of the medium
through the application of some appropriate mechanism. On the basis of
this study we can make an important conjecture that when the pressure
rise $\Delta p$ is negative, the couple stress effect creates
resistance to flow, whereby the velocity is reduced; however, the flow
is enhanced due to the couple stress effect in the case of free pumping
$\Delta p=0$\\

The present study reveals that it is possible to increase both pumping and pressure by
increasing the amplitude ratio and couple stress parameter and also by
reducing the permeability (Darcy number). The present study reveals
further that the velocity change is strongly dependent on $\phi$,
$\alpha$ and porosity. Lastly, the magnitudes of $\Delta p$,
$\phi$ and $\alpha$ affect significantly the nature of flow reversal.

\section*{Acknowledgments}
 The authors are thankful to the reviewers for their kind words of
 profound appreciation in respect of the quality and presentation of
 the work. One of the authors (Somnath Maiti) is thankful to the Council of
 Scientific and Industrial Research (CSIR), New Delhi for awarding him
 a senior research fellowship.

\end{document}